\input mtexsis
\input psfig
\paper 

\tenpoint

\def\mbf#1{\hbox{\mib$#1$}}

\def\d{ {\rm d } }

\referencelist

\reference{atha85} {Athanasoula, E. and Sellwood, J.} {\sl"Bi-symmetric
instabilities of Kuzmin-Toomre disks."} {\journal MNRAS;221,
213-232(1985)}\endreference
\reference{Bender} R. Bender, R. Saglia,  \&  O  Gerhard., 
{\journal MNRAS;269, 785-813(1994)}\endreference 
\reference{Bertin} {Bertin, G., }{\sl ``Linear stability of spherical collisionless stellar systems''}{\journal ApJ;434,94B(1994)}\endreference
\reference{collett} {Collett, J., (1995)}{ \it ``
Isocirculational mechanics in stellar systems
''}{ PhD, Cambridge}\endreference
\reference{DLB72} {Lynden-Bell, D. and  Kalnajs, A.} {\sl"On the
generating mechanism of spiral structures."} {\journal MNRAS;157,
1-30(1972)}\endreference
\reference{Clutton}{Clutton-Brok, M} {\journal MNRAS;157,
1-30(1972)}\endreference
\reference{earn} {Earn, D. and  Sellwood, J.} {\sl"The optimal N-body method
for stability studies of galaxies."} {\journal Astrophysical Journal v.451, p.533;,
0-0(1995)}\endreference
\reference{Earn} {Earn, D., (1993)}{}{ PhD}, Cambridge\endreference
\reference{hohl} Hohl F.   
{\journal ApJ;168, 343 (1971)}\endreference
\reference{Henon} {Henon, M.} {\sl"."} {Ann. d'Astrophys, 23 668 (1960).
}\endreference
\reference{Hunter} {Hunter, C.} {\sl"Instabilities of stellar discs."}
{Astrophysical disks ed. S.F. Dermott. J.H. Hunter and 
R.E. Wilson (New York New York Accademy of Sciences (1993)}\endreference
\reference{kalnajs71} {Kalnajs, A.} {\sl"Dynamics of Flat Galaxies. I."}
{\journal ApJ;166, 275-293(1971)}\endreference
\reference{kalnajs76} {Kalnajs, A.} {\sl"Dynamics of Flat Galaxies II,
Biorthonormal Surface density potential pairs for finite disks."}
{\journal ApJ;205, 745-761(1976)}\endreference
\reference{kalnajs77} {Kalnajs, A.} {\sl"Dynamics of Flat Galaxies VI,
The Integral Equation For Normal Modes in Matrix Form."} {\journal
ApJ;212, 637-644(1977)}\endreference
\reference{kalnajs78} {Kalnajs, A.}{(1978)}{} {{\it in} 
Berkhuisjen, E. Wielebinski, R. eds. Proc. IAU Symp. {\bf 77}, "Strutures and properties of 
nearby galaxies."} {p. 113} \endreference
\reference{Kuzmin} {Kuzmin, G. G.}{ Astr. Zh.  33 27 (1956)} \endreference
\reference{dlbo} {Lynden-Bell, D. and J. Ostriker} {\sl"On the
Stability of Differentially Rotating Bodies."} {\journal MNRAS;136,
293-310(1967)}\endreference
\reference{dlb79} {Lynden-Bell, D.} {\sl ``On a Mechanism that
Structures Galaxies.''} {\journal MNRAS;187,
101-107(1979)}\endreference
\reference{Miyamoto}{Miyamoto}{\sl ``A class of disk-like models 
for self-gravitating stellar systems''} \journal Astronomy \& Astrophysics;30,441-454(1974)\endreference
\reference{Pichon}{ Pichon, C. and Lynden-Bell, D.}
{\it ``Orbital instabilities in galaxies''}, {contribution to the
``Ecole de Physique des  Houches'' on: Transport phenomena
Jin Astrophysics, Plasma Physics, and Nuclear Physics}  \endreference
\reference{Pichon2} {Pichon, C. and Lynden-Bell, D.} {\it ``
The equilibria of flat and round disks''}, {{ {\it MNRAS.} (1996)
 {\bf 282 (4)},  1143-1158.}}  \endreference
\reference{Pichon3} {Pichon,~C. and Thi\'ebaut,~E. } {``Distribution
   functions for observed galactic  disks: a non parametric
 inversion'' }, {{ submitted to MNRAS  (1997)}}  \endreference
\reference{Qian} {Qian, E.} {\sl"Potential-Density Pairs for Flats
Discs."} {\journal Monthly Notices of the Royal Astronomical Society;257, 581-592(1992)}\endreference
\reference{sellwood} {Sellwood, J. and A. Wilkinson} {\sl"Dynamics of
barred galaxies."} {\journal Rep Prog Phys;56,
173-255(1993)}\endreference
\reference{sygnet} { Sygnet, J.F. \& Kandrup} 
{``A simple proof of dynamical stability for a class of spherical clusters''
}{\journal ApJ;276, 737(1984) }\endreference
\reference{Dejonghe} Vauterin,P. \& Dejonghe,H. ,
{\journal AA; 313, 465--477 (1996) }\endreference  
\reference{Zang} {Zang, T., (1976)}{ \it ``The stability of a Model
Galaxy''}{ PhD thesis, MIT}\endreference
\reference{Weinberg} {Weinberg, M. D., }{\journal ApJ;368, 66 (1991)}
{ }\endreference
\endreferencelist

\titlepage 
\title{Numerical linear stability analysis of round galactic disks}

%
%
\author
C. Pichon$^{1,2}$, \& R. C. Cannon$^{3,4}$ 
\centerline{$\bf 1$ Astronomisches Institut       
Universitaet Basel, Venusstrasse 7}
\centerline{CH-4102 Binningen Switzerland }
\centerline{$\bf 2$ CITA, McLennan Labs,University of Toronto,}
\centerline{60 St. George Street,Toronto,Ontario M5S 1A7.}
\centerline{$\bf 3$ Observatoire de Lyon, 9 Avenue Charles Andr\'e }
\centerline{  69 561 Saint Genis Laval, France.}
\centerline{ $\bf 4$University of Southampton, Bassett Crescent East}
\centerline{Southampton  SO16 7PX, United Kingdoom.}

\endauthor

\abstract

The  method   originally developed by  Kalnajs    for the  numerical  linear
stability analysis of round galactic  disks is implemented  in the regimes  of
non-analytic transformations  between position space and angle-action space,
and  of vanishing growth   rates.  This allows effectively any physically
plausible disk to be studied, rather than only those having analytic
transformations into angle-action space  which have formed the
primary focus of attention to date. The transformations are constructed
numerically  using orbit integrations in real  space, and the projections of
orbit radial   actions on  a   given potential density   basis  are Fourier
transformed  to  obtain  a  dispersion relation  in   matrix form.   Nyquist
diagrams are used to isolate modes  growing faster than  a given fraction of
the typical orbital   period and to  assess  how much  extra mass would   be
required to reduce the growth rate of the fastest mode  below this value. To
verify the implementation,  the fastest $m=2$  growth rates of the isochrone
and  the Kuzmin-Toomre disks  are recovered, and  the weaker $m=2$ modes are
computed.   The evolution of those  growth rates as  a  function of the halo
mass is also calculated,  and some $m=1$  modes are derived as illustration.
Algorithmic constraints   on   the   method's scope  are   assessed  and  its
application to observed disks is discussed.
\vskip 1cm

\raggedcenter{\ninepoint {\bf keywords}: Galaxies -- 
methods: analytical, numerical
--  stability, dynamics, disks  --  bars. }
\endraggedcenter

\toappear{ Monthly Notices of the Royal Astronomical Society}

\endtitlepage

\section{Introduction}

Theoretical  studies of the stability of   thin stellar disks provide useful
constraints on models of galactic  formation and dynamics.  Disk models  can
be excluded as unrealistic if they are found to be very unstable and the
results  of  galaxy formation  studies can  be clarified where  they lead to
models   close  to marginal   stability.  When  applied  to  observed disks,
stability analysis provides a unique tool to probe what fraction of the mass
is in luminous  form via the requirement  that there be enough extra matter
to make the observed distribution stable over a period comparable to the age
of the galaxy.

 Three main approaches have been adopted for disk stability analysis: direct
N-body simulations,  N-body simulations where  the bodies are smeared onto a
biorthonormal basis    thereby solving Poisson's    equation implicitly, and
linear modal analysis.   The first has been  widely used, initially  by Hohl
(1971)\cite{hohl}, and more recently,  {\it e.g.} by Athanasoula \& Sellwood
(1985)\cite{atha85}. It provides a flexible  tool of investigation which can
be  carried into  the non-linear  regime.   These simulations, however, only
provide insight  into instabilities in the  statistical sense, and generally
probe poorly the marginal stability regime.  This drawback has recently been
addressed by Earn  \& Sellwood  (1995)\cite{earn}  who developed the  second
approach  of solving Poisson's equation  through a biorthonormal basis.  The
stability of stellar  systems has also  been explored  for spherical systems
via   a  global      ``energy   principle''  (see     Sygnet     \& Kandrup.
(1984)\cite{sygnet}) but this approach has not been successfully implemented
for disks because of resonances (Lynden-Bell \& Ostriker (1967)\cite{dlbo}).
Moreover, the  method  usually  provides stability statements  which  are of
little practical  use, since no time scale  for the astrophysically relevant
growth rates is available.

The third   method, linear  modal  analysis (Kalnajs (1977)\cite{kalnajs77},
Zang (1979)\cite{Zang} Hunter (1992)\cite{Hunter}),  is used here.  In spite
of the  obvious  limitation  to  perturbations  of small amplitudes  it  has
several potential   advantages.  In this  approach  the integro-differential
equation  resulting   from self-consistent solutions   of  the Boltzmann and
Poisson  equations  is recast  into   a  non-linear eigen-problem using   an
appropriately defined orthonormal  basis.  The method, pioneered  by Kalnajs
(1971-77)\cite{kalnajs71}\cite{kalnajs76}  \cite{kalnajs77} on the    linear
stability of galactic   disks, involves  the  formulation of  the  dynamical
perturbed equations  in angle-action  coordinates  and the restriction  to a
(finite) set of perturbed densities which  diagonalise Poisson's equation in
position   space.    Kalnajs' original  work\cite{kalnajs71}  concerned  the
stability  of the isochrone disk for  which explicit transformations between
angle-action   and  position-velocity   coordinates  exist.  Zang 
(1979)\cite{Zang}
studied the  self-similar Mestel disks in  the same way.  More recently, the
method  was  implemented   by   Hunter  (1992)\cite{Hunter}    (via repeated
evaluation of  elliptical integrals)  for  the infinite Kuzmin-Toomre  disk.
The approach of these  authors involved calculating explicit transformations
between angle-action variables  and   position-velocity variables and    was
therefore restricted to a very limited set of simple analytic disks.
More recently,   the relevant  angle   action
integrals  were computed   by quadrature   by Weinberg (1991)\cite{Weinberg},
and Bertin (1994)\cite{Bertin} while
studying the  stability of spherical  systems  for which  the Hamiltonian is
separable. Finally Vauterin and Dejonghe (1996)\cite{Dejonghe}
carried out a similar analysis for disks while performing the whole 
investigation in position space.

Here, the first step of  mapping the distribution function into angle-action
space  is     implemented numerically   by   calculating     the appropriate
transformations  from the results of  integrating unperturbed  orbits in the
mean  field of the  galaxy.  This  then allows  considerable  freedom in the
initial equilibrium to be studied, including the prospect of applying linear
stability    analysis to   distribution   functions   recovered for observed
galaxies. This method is not  restricted to integrable potentials and  could
be generalised to 3D. It can in particular be implemented for {\sl measured}
distribution functions, where the  potential  is deduced from  the  rotation
curve.

In section~2 Kalnajs' matrix method is recovered following an indirect route
corresponding to a rewriting of Boltzmann's and Poisson's equations directly
in action space.   The  stability criteria  are then reformulated  in a form
suitable   for  numerical evaluation of   the  matrix elements in Section~3.
Details of the  numerical method are presented and  its range of validity is
discussed.  Section~4 contains results of calculations for the isochrone and
the Kuzmin-Toomre disks.  These are in agreement with those found by Kalnajs
(1978)\cite{kalnajs78}, Earn  \& Sellwood (1995)\cite{earn}, Athanassoula \&
Sellwood  (1985)\cite{atha85} and Hunter (1992)\cite{Hunter}.  Prospects and
applications to observed data as  probes of dark  matter are sketched in the
last section.

\section{Linear stability analysis revisited}

Linear   stability analysis concerns   the  dynamical evolution of initially
small perturbations in their  own self-consistent field. The derivation (see
{\it eg}  Kalnajs (1977)\cite{kalnajs77}, Sellwood \& Wilkinson (1993)\cite{sellwood}) of
the resulting   integro-differential equation is  sketched here  putting the
emphasis on a  coherent description in action space.  This analysis is to be
contrasted with that of Vauterin \& Dejonghe (1996)\cite{Dejonghe} who chose
to implement  stability analysis  for   disks in position  space alone.  The
resulting dispersion relation,   \Eq{C16}, is fully equivalent  to  Kalnajs'
Eq.~K-15, but the intermediate integral equation, \Eq{4-27}, also provides a
connection with  orbital stability analysis (Lynden-Bell (1979)\cite{dlb79},
Pichon    \&      Lynden-Bell   (1992)\cite{Pichon}   and    Collett
(1995)\cite{collett} ).  \nl

\subsection{The integral equation}

The Boltzmann-Vlasov equation, 
$$\partial F/\partial t  + [H,F] = 0 \, ,\EQN
 4-1$$  
governs  the dynamical   evolution of  an ensemble  of collisionless
 stars.  Here $H$ is the Hamiltonian for the motion  of one star, $F$ is the
 mass-weighted distribution function in phase space,  and the square bracket
 denotes the Poisson bracket.
Writing $F = F_0 + f$, and $\Psi = \psi_0 + \psi$, 
where $F_0,\psi_0$ are the  unperturbed distribution function 
and potential respectively, and linearising 
\Eq{4-1} in $f$ and $\psi$, the perturbed distribution function 
and potential, yields
$$\partial f/\partial t + [H_0, f] - [\psi, F_0] = 0 \, .\EQN 4-3$$
  According to Jeans' theorem, the unperturbed equation, 
$[H_0, F_0] = 0$, is solved if $F_0$ is any function only of the 
specific energy,
$\varepsilon$, and the specific angular momentum, $h$. 

Following Lynden-Bell and Kalnajs (1972)\cite{DLB72}, angle and action
variables of the unperturbed Hamiltonian $H_0$ are chosen here as
canonical coordinates in phase-space.  The unperturbed
Hamilton equations are quite trivial in these variables, which makes
them suitable for perturbation theory in order to study quasi-resonant
orbits. 
The actions are defined in terms of the  polar coordinates, $(R,\theta)$, by
${\bf J} = (J_R, J_\theta)$, where 
$$J_R = {( 2\pi)^{-1}} \oint
[{\dot R}] \d R\, , \quad {\rm and } \quad J_\theta = h = R^2 
{\dot\theta} \, . \EQN$$ Here
$[{\dot R}]$ is a function of the radius, $R$, the specific energy, 
$\varepsilon$,
and the specific angular momentum, $h$, given by $[{\dot R}] =
\sqrt{2\varepsilon + 2\psi_0 (R) - h^2/R^2}$.
The angle between 
apocentres is
$\Theta = \oint h/R^2 [{\dot R}]^{-1} \, \d R$.
For a radial period $T_R$, 
and  a pair of azimuthal and radial epicyclic frequencies
$\Omega$, and $\kappa$, given by  
$ \Omega = \Theta / T_R $ and $ \kappa = 2 \pi / T_R  $
the phase-angles
conjugate to $J_R$ and $h$ are  ${\mbf
\varphi} = (\varphi_R, \varphi_\theta)$, where 
$$\varphi_R = \kappa \int^R [{\dot
R}]^{-1} \, \d R \, ,
\quad {\rm and } \quad \varphi_\theta =
 \theta + \int^R (h/R^2 - \Omega)[{\dot R}]^{-1}\, \d R\, . \EQN 4-7 
$$

  The stationary unperturbed Boltzmann equation \Eq{4-1} is solved by 
any distribution function of the form $F = F_0 
({\bf J})$, since $[H_0, {\bf J}] = 0$.

For growing instabilities,  $f$ and $\psi$ are both taken
to be 
proportional to $e^{-i\omega t}$, $\omega$ having a positive imaginary part. 
When expanded
 in Fourier series with respect to the angles $\mbf \varphi $,
\Eq{4-3} becomes after simple algebra (Kalnajs 1977\cite{kalnajs77})
$$f_{\bf m} = {m^{-1} {\bf m}\cdot\partial F_0/\partial {\bf J}\over 
\Omega_\ell - \Omega_p} \, \psi_{\bf m} \, ,\EQN 4-14$$
where
 ${\bf m}$ is a integer vector with components $(\ell, m)$,  
 $\Omega_\ell = m^{-1} {\bf m}\cdot {\bf \Omega} = \Omega +
\ell\kappa/m$ ,
   ${\bf \Omega} $ stands for $(\kappa, \Omega)$,
and $\Omega_p =  \omega/m$.
Here
 $\psi_{\bf m}$ and $f_{\bf m}$ are  the Fourier 
transforms of $\psi$ and $f$ with respect to 
$(\varphi_R,\varphi_\theta)$; for instance
$$ \psi_{\bf m}({\bf J})= {1 \over 4\pi^2} \int 
\psi(R,\theta) \exp \left[\, -i( {\bf m}\cdot {\mbf\varphi}) \right]\, 
d^2\varphi \, . \EQN fourier$$

Poisson's integral relates the potential, $\psi$, to the density 
perturbation:
$$\psi(R',\theta') = { 4 \pi G} \int {f({\bf R},{\bf v})\over |{\bf R}- {\bf R}'|  }\,\d R\,\d\theta \,
 \d v_R \, \d v_\theta \, .\EQN 4-21 $$
 This equation may be rewritten in order to make explicit the
contribution from the interaction of orbits. Here again angle-action
variables are useful, as a given unperturbed orbit is entirely
specified by its actions. It is therefore straightforward to identify
in Poisson's integral the contribution corresponding to the
interaction of given orbits.  Re-expressing this equation in terms of
angles and actions $({\mbf \varphi},{\bf J})$, using Parseval's theorem
and taking its Fourier transform with respect to ${\mbf \varphi }$
leads to:
$$\psi_{\bf m} ({\bf J}) = 4 \pi G \sum_{\bf m'} \int f_{\bf m'} ({\bf 
J}') A_{\bf mm'} ({\bf J},{\bf J'})\, \d^2J' \, , \EQN 4-22$$
where $\psi_{\bf m}$ and $f_{\bf m}$ are given by \Eq{fourier} and
$$A_{\bf mm'} = {1\over (2\pi)^4} \int {\exp\, i({\bf m}'\cdot 
{\mbf\varphi}' - {\bf m}\cdot {\mbf\varphi})\over |{\bf R} - {\bf R}'|} 
\d^2\varphi'\, \d^2\varphi \, .\EQN 4-23$$
The double  sum  in \Eq{4-22} extends   in  both $\ell$ and   $m$ from minus
infinity to infinity where the radii $ {\bf  R} ({\mbf\varphi},{\bf J})$ and
$  {\bf R} ({\mbf\varphi}', J')$  are  re-expressed  as  functions of  these
variables.  Now  $|{\bf R}  -  {\bf R}'|$  depends on  $\varphi_\theta$  and
$\varphi'_\theta$   in the  combination  $\varphi'_\theta  -  \varphi_\theta
\equiv  \Delta\varphi$    only.  As    $|\partial    \left(\varphi'_\theta\,
\varphi_\theta \right) / \partial
\left( \Delta\varphi\, 
 \varphi_\theta \right)| = 1$, 
the order of integration  in \Eq{4-23} may then be reversed, doing the $\varphi_\theta$ 
integration with $\Delta\varphi$ fixed.  
This yields $2\pi\delta_{mm'}$, so 
${\bf m}'$ becomes $(\ell', m)$ in the surviving terms. This  gives for \Eq{4-23}
$$A_{\bf mm'} = { 2 \pi \delta_{mm'}\over (2\pi)^4} \int {\exp\, 
i(m\Delta\varphi - \ell'\varphi'_R + \ell\varphi_R) \over |{\bf R} - 
{\bf R}'|} \d\Delta\varphi \, \d\varphi'_R\, \d\varphi_R \, .\EQN mention
$$ Each $m$ mode therefore evolves independently.  The dependence on
$m$ will be implicit from now on. So, for example,  $\psi_{\ell}$ is the Fourier 
transform of $\psi(R,\theta)$ with respect to both $\varphi_R$ and $\varphi_\theta$.
\nl
Putting \Eq{4-14} into \Eq{4-22}
 leads to the integral equation
$$\psi_{\ell_1}({\bf J}_1) = 4 \pi G \sum_{\ell_2} \,
\int \,
 A_{ \ell_1 \ell_2}( {\bf J }_{ 1}, {\bf J}_{ 2}) \, 
{{m_2}^{-1} {\bf m_2}\cdot\partial F_0/\partial {\bf J}_2\over 
\Omega_{\ell_2} - \Omega_p}
\,\psi_{\ell_2}({\bf J}_2) \,
\d^2{\bf J}_2 \, .\EQN 4-27$$
This equation is the integral equation for the linear growing mode 
with an m-fold symmetry of a thin disk. 
\Eq{4-27}  was later approximated by Pichon \& Lynden-Bell (1992)\cite{Pichon}
Collett (1995)\cite{collett} Lynden-Bell (1995)
to analyse the growth of the perturbation in terms of a Landau instability.

\subsection{The dispersion relation}
The perturbed distribution function and potential are now expanded 
over a potential-distribution basis $\{f^{(n)}\}_{n}$, $\{ 
\psi^{(n)}\}_{n}$ as 
$$ f(R,V) = \sum_{n} a_{n}  f^{(n)}(R,V)\,, \quad {\rm and } \quad 
\psi(R) = \sum_{n} a_{n}  \psi^{(n)}(R)  \, , \EQN expansion $$
where  the  basis is assumed to satisfy Poisson's equation 
\Eq{4-21}, which when written in action space has Fourier components 
obeying (following
\Eq{4-22} and given \Eq{mention}):
$$\psi_{\ell}^{(n)} ({\bf J}) =  4 \pi G \sum_{\ell'} \int 
f_{\ell'}^{(n)} ({\bf J}') A_{ \ell \ell'} ({\bf J},{\bf J'})\, \d^2J' \, . 
\EQN 4-222$$ 
For basis functions  scaling like $\exp(i m' 
\theta)$ (preserving the axial symmetry 
described above), \Eq{fourier} applied to, say, 
$\psi^{(n)}(R) \exp(i m' \theta)$ gives for 
the Fourier mode $\ell$ : 
$$\psi _\ell ^{(n)}({\bf J})={1 \over {2\pi }}\int_0^{2\pi 
} {}e^{-i\ell
\varphi _R}\psi ^{(n)}(R\left[ {\varphi _R} \right])e^{im\delta 
\theta(\varphi _R)}\d\varphi _R \, ,  \EQN fourierII$$
where \Eq{4-7} has been used to define the relative azimuthal increment
$\delta \theta (\varphi  _R) \equiv \varphi_\theta -\theta$. 

 Expanding $\psi$ over this basis (using the {\sl same} expansion for all 
 $\ell$s) according to \Eq{expansion},  inserting this expansion into 
 \Eq{4-27},  multiplying by  $ f_{\ell}^{(p) *} ({\bf 
J})$,  integrating over $\bf J$ and summing over $\ell$  yields
 $$ \sum_{n}  a_n (\sum_{\ell_1}  \int \psi_{\ell_1}^{(n)}({\bf J}_1) 
 f_{\ell_1}^{(p) *}({\bf J}_1)
 \d^2{\bf J}_1 ) = $$ 
 $$\sum_{n'}  a_{n'}  (
  4 \pi G \sum_{\ell_2,\ell_1}  \,
\int \, \int \, f_{\ell_1}^{(p) *}({\bf J}_1)
 A_{  \ell_1 \ell_2}( {\bf J }_{ 1}, {\bf J}_{ 2}) \, 
{{m_2}^{-1} {\bf m_2}\cdot\partial F_0/\partial {\bf J}_2\over 
\Omega_{\ell_2} - \Omega_p}
\,\psi_{\ell_2}^{(n')}({\bf J}_2) \,
 \d^2{\bf J}_2 \, \d^2{\bf J}_1 ) \, .\EQN 4-277$$
Since $f_{\ell_1}^{(p)}$ belongs to the basis, it satisfies \Eq{4-222} 
and consequently
$$ \sum_{n}  a_n   ( \sum_{\ell_1} \int \psi_{\ell_1}^{(n)}({\bf J}_1) f_{\ell_1}^{(p) *}({\bf J}_1)
 d^2{\bf J}_1 ) = \sum_{n'}  a_{n'}  (
    \sum_{\ell_2}  \,
\int \,  \psi_{\ell_2}^{(p) *}({\bf J}_2)\,  \, 
{{m_2}^{-1} {\bf m_2}\cdot\partial F_0/\partial {\bf J}_2\over 
\Omega_{\ell_2} - \Omega_p}
\,\psi_{\ell_2}^{(n')}({\bf J}_2) \,
 \d^2{\bf J}_2 ) \, .\EQN 4-2777$$
 
 Requiring that \Eq{4-2777} has non trivial solutions in 
 $\{a_n\}$ leads to the dispersion relation 
$$ D(\omega) =  \det |{\bf \Lambda} - {\bf M}(\omega) | = 0 \, , \EQN C16$$
 where the matrix $\bf M$ is defined in terms of its components $(n',p)$ by the bracket on the {\it r.h.s.}
 of \Eq{4-2777} while the  matrix $\bf \Lambda$ corresponds to the 
 identity if the basis $(\psi_n, f_n)$ is bi-orthonormal, or to a matrix 
 with components
 components $(n,p)$ by the bracket on the {\it l.h.s.}
 of \Eq{4-2777}
 otherwise. Equations \Ep{C16} was first derived in this context by Kalnajs
(1977\cite{kalnajs77}).
In order to approximate the behaviour of a halo,  a supplementary parameter 
$q \in ]0,1]$ is introduced
$$ D_q(\omega) =  \det |{\bf \Lambda} -  q {\bf M}(\omega) | = 0 \, , \EQN C161$$ 
as discussed below.

\subsection{Unstable modes \& Nyquist diagrams } 
The dispersion relations \Ep{C161} give, for each $m$, the criterion for the 
existence of exponentially growing unstable modes of the form $\exp( -i 
\omega t+ i m \theta)$. They are functions of a free complex 
parameter $\omega$
 corresponding in its real part to m times the pattern speed of the growing
mode,   $\Omega_p$, and in  its imaginary   part to the growth  rate  of the
perturbation.  The search  for the growing  modes is  greatly facilitated by
the use of   Nyquist diagrams. Consider  the  complex $\omega$  plane  and a
contour that traverses along the real axis and then closes around the circle
at $\infty$ with ${\cal I}m  (\omega)$ positive.  The determinant $D_{q}$ is
a  continuous  function of the   complex variable  $\omega$, so as  $\omega$
traces  out the closed contour,  $D_{q}(\omega)$ traces out a closed contour
in the complex $D_{q}$ plane.  If  the $D_{q}(\omega)$ contour encircles the
origin, then there is a zero of the determinant  inside the
original contour and 
the system is  unstable. In fact  a simple argument of  complex
analysis shows that the number of loops around the origin corresponds to the
number of poles above the line ${\cal 
I}m  (\omega) \equiv \eta = {\rm Constant}$.  An intuitive picture of
how  the criterion operates   on \Eq{C16}  is  provided  with the  following
thought  experiment.  Imagine  turning up  the   strength of $q  G$  for the
disk ({\it    i.e.}   turn on self-gravity,      which is physically
equivalent to  allowing for the  gravitational interaction to play its role;
alternatively vary $q$, the ratio of relative  halo support).  Starting with
small  $q  G$, all the $M^{n   p}$ are small, so  for   all $\omega$'s the
determinant $D_{q}(\omega)$ remains on a small contour close to $ \det |{\bf
\Lambda }|$.  As $q G$ is increased to its full value, either the  $D_{q}$ 
contour  passes through the origin to  give a marginal instability  or it does not.
If  it does  not,  then  continuous change of   $q  G$ does  not modify  the
stability, and therefore the self-gravitating system  has the same stability
as in the zero $q G$ case:  it is stable.   If, however, it crosses,
and remains circling  the  origin then it has  passed  beyond the marginally
stable case  and is  unstable.    This picture is   illustrated in  the next
sections where   the  image of  the  complex line   $ m  \Omega_p+i\eta   $,
$\Omega_p \in ]-\infty, \infty[$, is plotted for various values of 
$\eta$.

\section{Numerical implementation}

\subsection{Method}

Stability  analysis, as  implemented in  the  previous  section, allows  the
treatment of a wide range of potentials and distribution functions which, in
general, will not  afford the explicit transformations between angle--action
variables and velocity--position variables.

Spline interpolation for tabulated potentials  and distribution functions is
used  to compute the transformation of  variables by numerically integrating
orbits in the given potential.  The whole calculation is performed on a grid
of points in  $(R_0, V_0)$ space where  $R_0$ is the  radius at  apocentre, and
$V_0$ the  corresponding tangential velocity.   As may be seen  from  figure
\Fig{nrnv-rmaxplot} in the next  section, a $40  \times 40$ grid is adequate
for the disks considered here.

For each point in  this grid, three orbits are  calculated, one starting  at
the grid-point itself, and  the  others at small  deviations  in $h_0 =  R_0
V_0$, and along $h_0 =  {\rm constant}$.  The  angles and actions for  these
orbits are used later to calculate numerical derivatives of the distribution
function. A fourth order Runge-Kutta  integration scheme is employed for the
orbits, stopping   at the first  pericentre  for all but the  most eccentric
orbits where more than one oscillation is needed to give sufficient accuracy
in the actions.  Further points on the orbits corresponding to grid vertices
are then   calculated at  exact  subintervals of  the  orbital  period.  The
Fourier components $\psi^n_l$ of a particular basis element along an orbit
\Eq{fourierII} are then given by:
$$\psi _\ell ^{(n)}(R_0,V_0)={1 \over {T_R}}\int_0^{T_R} {}e^{-2 \pi i \ell 
{t / {T_R}}_{}}\, \,\psi ^{(n)}(R\left[ t \right])e^{im\delta \theta (t)}\d t 
\, ,  \EQN fourier2$$
where  $t= \varphi_R  /\kappa $ and
$$  \delta \theta (\varphi _R)  = 
\kappa^{-1}\int^{\varphi_R} { h  \over
 5R^2 } \, {d
 \varphi_R } -\kappa^{-1} { \varphi_R \over 2 \pi} 
 \oint { h \over R^2 }  \, {\d \varphi_R } \,=
\theta(t)  - \langle {\dot \theta(t)} \rangle t  \, . \EQN   $$
Numerically, this  simply  corresponds to taking  a  discrete Fourier 
transform (DFT) over the  re-sampled
points.    The symmetry of  an  orbit means that   although  the argument is
complex, the result is purely real.

\midfigure{NyquistPic}
\vskip 14cm
\special{voffset=0 hoffset=0 hscale=80 vscale=120 psfile=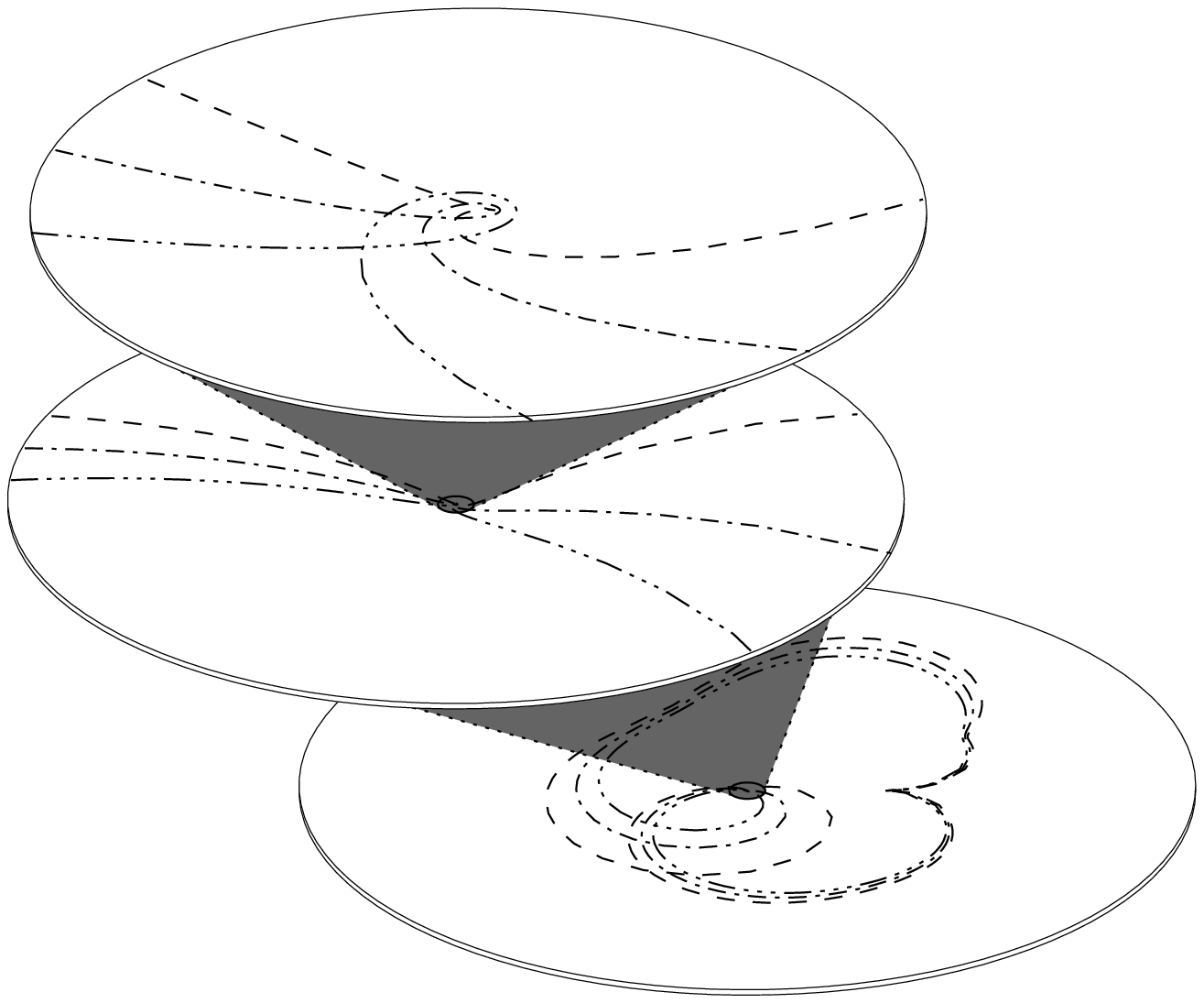}
\Caption
{Nyquist diagram  for the fastest  growing mode of  an isochrone/9 model for
 growth rates $\eta = 0.12$ (dashed curve),  $\eta = 0.14$ (dot-dashed curve)
and  $\eta  = 0.16$ (dot-long-dashed   curve).  
The magnification shows that  the
 first two  curves  do enclose   the  origin whereas   the third does   not,
 indicating that the true growth rate is between $0.14$ and $0.16$.  }
\endCaption
\endfigure

The matrix element defined in \Eq{4-2777} is rewritten in 
terms of $(R_0,V_0)$ and truncated in $\ell$: 
$$M_{}^{(n)(n')}(\omega)=\sum\limits_{\ell =-L}^{\ell =L} 
{}\int\!\!\!\int {}\psi _\ell ^{(n)}{}^*{{
\left( {{{\partial \,F} / {\partial \,h}}\cdot m+{{\partial \,F}
 / {\partial \,J}}\cdot \ell } \right)} \over 
  { {\Omega 
\cdot m+\kappa \cdot \ell -m \Omega_p }-i\eta }}\psi _\ell ^{(n')}\left|
 {{{\partial  \,J\partial h} \over {\partial \,R_0\partial V_0}}} 
  \right|\kern 1pt \d R_0\kern 1pt \d V_0  \EQN  Mn1n2$$
where the $\psi _\ell ^{(n')}$ are now surfaces in $(R_0, V_0)$ space
given by the $\ell$'th order term of the DFT of the orbit at $R_0, V_0$.

The inclusion of retrograde stars may be effected transparently by 
specifying a grid in $R_0, V_0$ which includes negative $V_0$ or,
 more efficiently, by noting that 
$$\EQNalign{\psi _\ell ^{(n)}(-h) &={1 \over {\pi}}\int_0^{\pi} \cos({ \ell 
{\varphi}_{R}(-h)-m \delta \varphi(-h)})\, \,\psi ^{(n)}(R\left[ \varphi_R(-h)\right])\d\varphi_{R}
\, ,  \EQN fourier3;a \cr 
 &={1 \over {\pi}}\int_0^{\pi} \cos({ \ell 
{\varphi}_{R}(h)+m \delta \varphi(h)})\, \,\psi ^{(n)}(R\left[ 
\varphi_R(h)\right])\d\varphi_{R}
\, ,  \EQN fourier3;b  \cr
&=\psi _{-\ell}^{(n)}(h)
\, ,  \EQN fourier3;c  \cr }$$
so with the appropriate sign switching the same set of orbits and
Fourier modes may be used as for prograde stars.

Having calculated the derivatives of the  distribution function with respect
to $J$ and $h$  by finite  differences across  the slightly displaced  orbits
described earlier and the Jacobian $ {{{\partial ( J\, h)} / {\partial
(R_0\, V_0)}}}$ in the same manner, the problem of computing the matrix
elements of equation  \Eq{Mn1n2} reduces to one  of integrating quotients of
surfaces on  the $R_0, V_0$ grid.   It is worth  noting that, although it is
the steps of  the calculation up to  this point which provide the generality
of the  implementation, their computational cost is  relatively slight
amounting in total, for example, to about a minute  on a workstation with a
specFp92 of 100.

For vanishing  growth-rates,  the  integrand   in  equation \Eq{Mn1n2} is   a
quotient  of surfaces  with  a real numerator  but  a denominator  which may
vanish  along  a line within the  region  of integration.  The integral then
exists only as  a principal value integral.  Rather  than attempting a brute
force discretisation we   approximate  both  numerator  and denominator   by
continuous piecewise flat  surfaces and employ  analytic  formulae, or their
power series expansions, for each flat subsection.  Each square of the $R_0,
V_0$  grid  is treated as  two  triangles.  According to the  magnitudes and
slopes of  the two surfaces  over a given  triangle  there are six different
approximations to be used for  each  component of  the integral.  For  modes
with finite growth-rates,   the integrand in   equation  \Eq{Mn1n2} is of   a
quotient of surfaces  with a  real  numerator but a complex  denominator and
this  prescription  still holds.   The  extra  generality afforded yields  a
solution which handles transparently  the marginal stability case, where the
resonances can be infinitely sharp.

\midfigure{NyquistDiagram1}
\vskip 6.5cm
\special{voffset=-20 hoffset=0 hscale=40 vscale=50 psfile=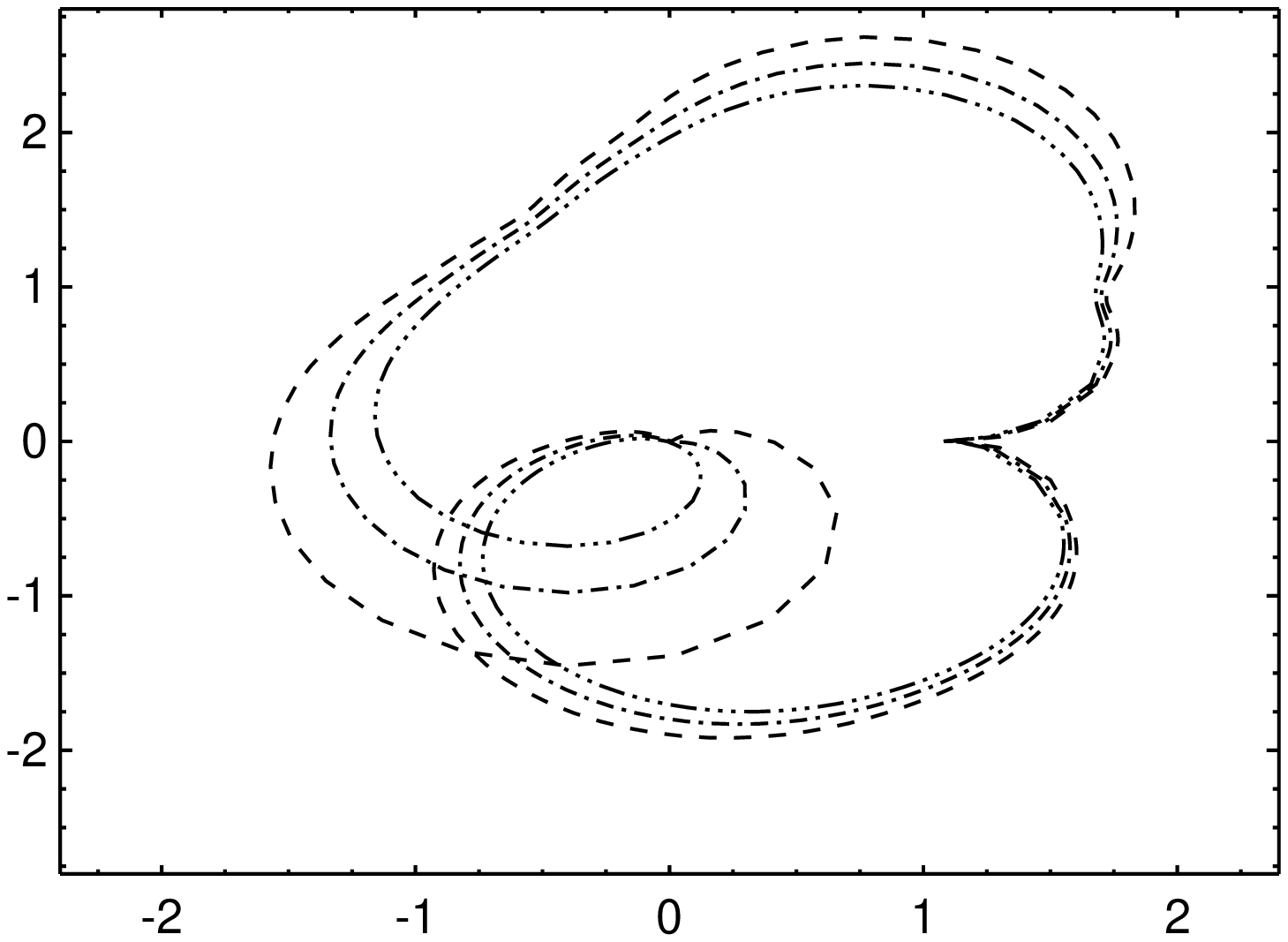}
\special{voffset=-20 hoffset=220 hscale=40 vscale=50 psfile=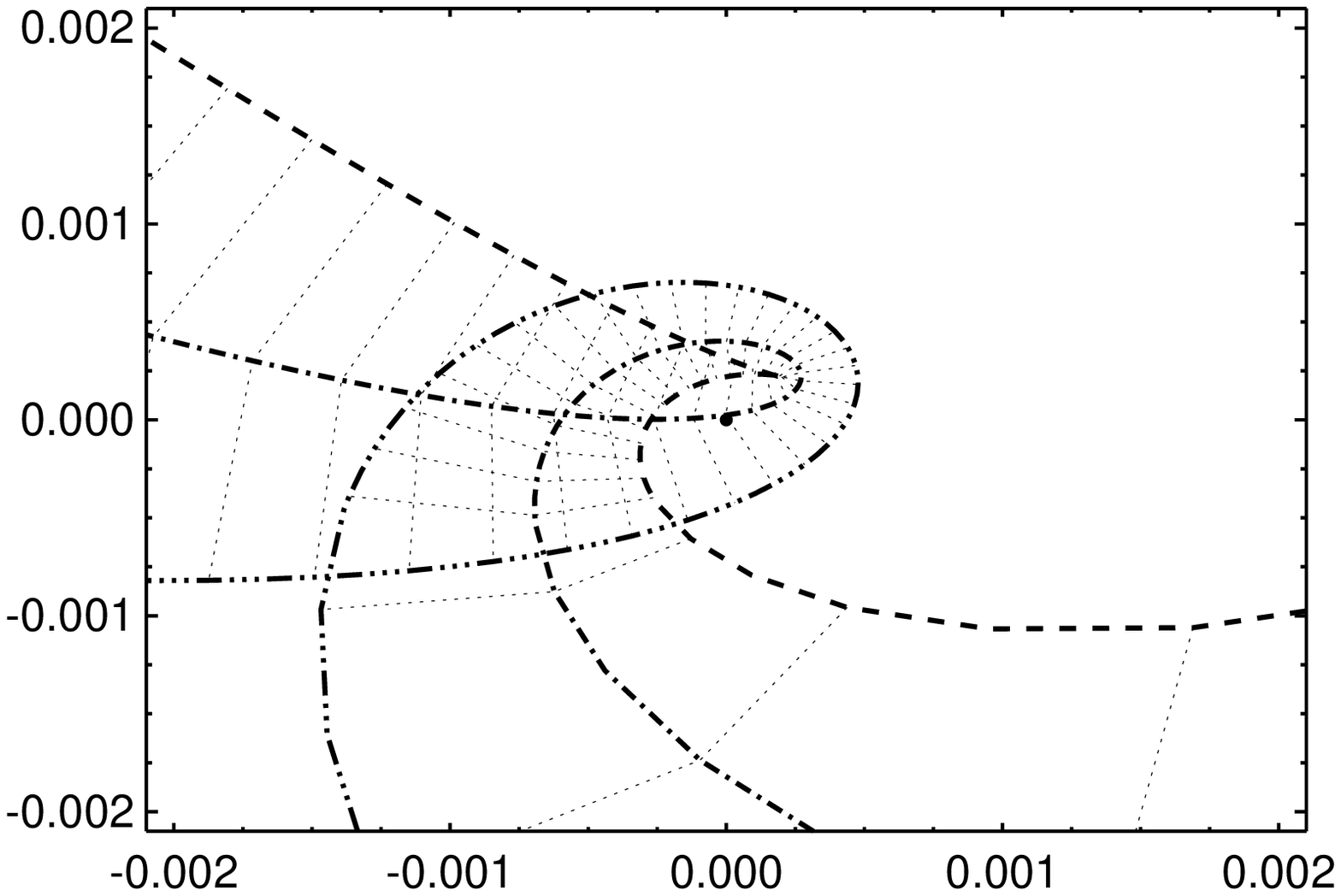}
\Caption
{Nyquist diagram for the fastest growing  mode  of
 an isochrone/9 disk with curves labelled as on \Fig{NyquistPic}.
The dotted lines join points of the same $\omega$.}
\endCaption
\endfigure

\midfigure{NyquistDiagram2}
\vskip 6.5cm
\special{voffset=-20 hoffset=0 hscale=40 vscale=50 psfile=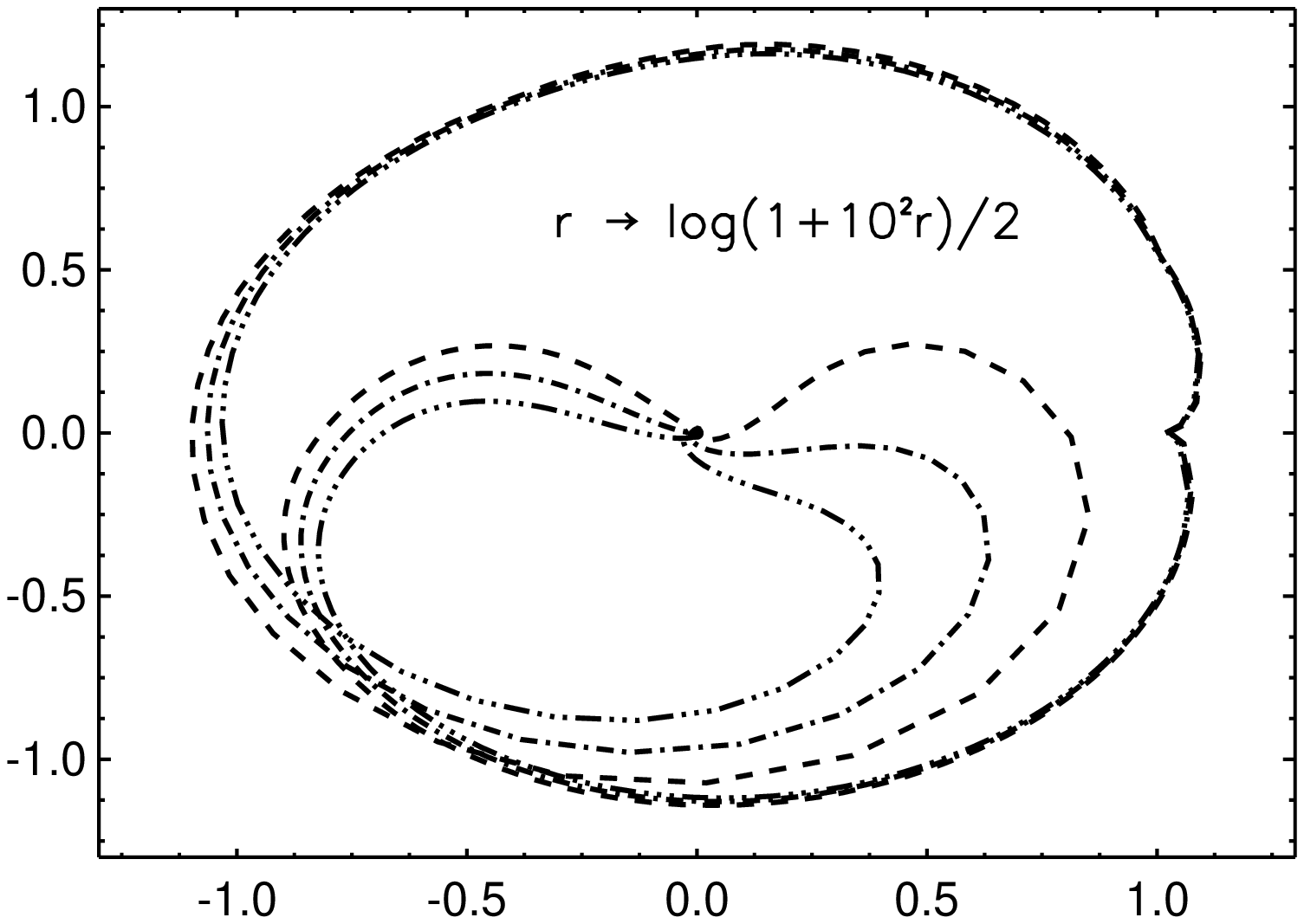}
\special{voffset=-20 hoffset=220 hscale=40 vscale=50 psfile=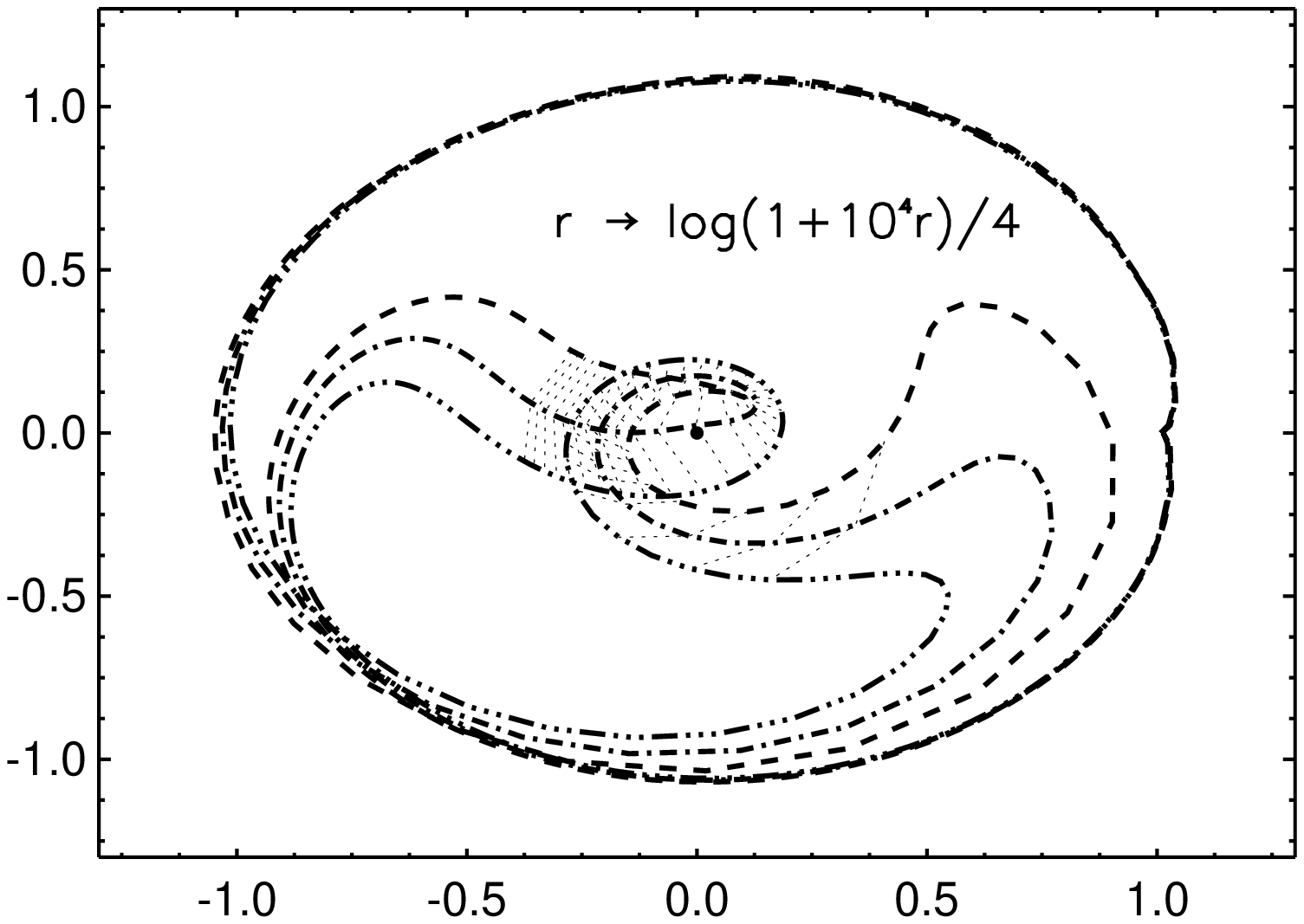}
\Caption
{Nyquist diagram for the fastest growing mode of an isochrone/9
as in \Fig{NyquistPic},\Fig{NyquistDiagram1}.  A  logarithmic scaling has been
applied near the origin mapping $r \rightarrow \log_{10} (1 + 10^\alpha r) /
\alpha$ allowing the full topology to be seen more easily.}
\endCaption
\endfigure

The integration  and summation then yields,  for each $\Omega_P$ and $\eta$,
an $n \times n$ matrix $M$. Marginal  stability corresponds to the vanishing
of ${\rm  det} | {\mbf  I} -{ \mbf M}|$.  As described in  section 2.3 it is
convenient to use Nyquist  diagrams to locate  the critical points: matrices
corresponding to a hundred or so values of $\omega$ are calculated for lines
of constant  $\eta$. Providing the  sampling is  good enough,  the number of
times the line ${\rm det} | {\mbf  I} -{ \mbf M}| $ encircles the origin in  the complex plane gives
the number of zeroes   above the initial line $m   \Omega_p + i  \eta$.   An
example  of nyquist diagrams  for   the  isochrone/9  disk  (section~4)   is
illustrated in
\Fig{NyquistPic} and  shown  more quantitatively  in
\Fig{NyquistDiagram1} and 
\Fig{NyquistDiagram2}. The 3-dots-dashed curve  corresponds to $\eta = 0.16$
and does not encircle  the origin, (or  rather it  loops once  clockwise and
once   anticlockwise) whereas  the  other two    curves ($\eta  =  0.12$ and
$\eta=0.14$) do,  indicating that the growth rate  for the  fastest growing
bisymmetric instability of this disk is between $0.14$ and $0.16$.

This method  is easy to apply to  verify calculations where  the growth rate
and pattern speed are  already known, but  becomes more time-consuming when
treating disks   with unknown properties.   Many automatic  schemes  can  be
envisaged for locating zeroes in  the complex plane  but since computing the
matrices is the slowest  part of the  calculation, we have found it
most efficient to manage the search by hand with  an  interactive tool
for  examining  Nyquist
diagrams for different values of $q$  (section 2.3). With a little experience
the zeroes can be efficiently located even from poorly sampled diagrams with
spurious  loops which  would  easily  lead to  confusion  in  automated
procedures.

\subsection{ Validation}

There are  two distinct  types of errors  to be  assessed in this  analysis:
numerical   errors  arising from the discretization    of integrals and from
machine rounding; and truncation errors relating to  the use of small finite
bases, limited Fourier  expansions  and the  calculation  of only the  inner
parts of infinite disks.

While in principle constraining the first type  of errors is purely routine,
it is worth noting  that both the bases  considered here require the  use of
arbitrary  precision languages (such as {\tt   bc}) for their evaluation. As
noted by  Earn \& Sellwood (1995)\cite{earn}, Kalnajs' basis   requires the use of about  fifty
figures precision  to  get the  first thirty basis  elements to five
figures. The  problem, however,  lies  only in evaluating the   polynomial
terms because  the coefficients are large and induce substantial cancelations.
Qian's basis (1992)\cite{Qian}   not  only requires extended precision    to
evaluate a single element but it is also extremely  close to being singular.

Besides evaluation of   the basis elements,  the  only other part  requiring
special care is  in the numerical  derivatives of  the distribution function
with respect to  $J$ and $h$ where  a tight compromise  must be made between
the desire for a small interval to give an accurate derivative, and the loss
of significance in the differences between angles and actions calculated for
very close orbits.  Nevertheless,  a straightforward Runge-Kutta scheme  for
the orbits remains adequate in all the cases studied here.

The second class of errors -- those arising from the truncation of equations
\Eq{C16} --  are best assessed by  simply recomputing the same quantity with
more basis functions, more Fourier harmonics, or a  larger disk.  The results of
these  tests     for     all  the truncated   quantities      can   be  seen
in \Fig{nrnv-rmaxplot} and \Fig{k-lplot}. 
   All  these  results   are   for  the
isochrone/9 model first studied  by  Kalnajs (1976)\cite{kalnajs76} with the  $n=7$
series of  Kalnajs' basis functions, which is  further discussed in the next
section.  Since  the   calculation  of the  matrix  ${\bf   M}(\omega)$ in
\Eq{C161} is the  most computationally intensive operation, the self-gravity
parameter $q$ is taken  instead of the  growth rate $\eta$ as a  convergence
criterion. The $q$ plotted in the figures is that which satisfies \Eq{C161}.
It is found by bisection with respect  to the winding  number of the Nyquist
diagram.  A value $q=1$ for the expected exact growth  rate
(given by Kalnajs (1978)\cite{kalnajs78}) is the asymptote
towards which the calculation should  converge when the appropriate range of
parameters have been found.

\midfigure{nrnv-rmaxplot}
\vskip 6.5cm
\special{voffset=0 hoffset=20 hscale=40 vscale=50 psfile=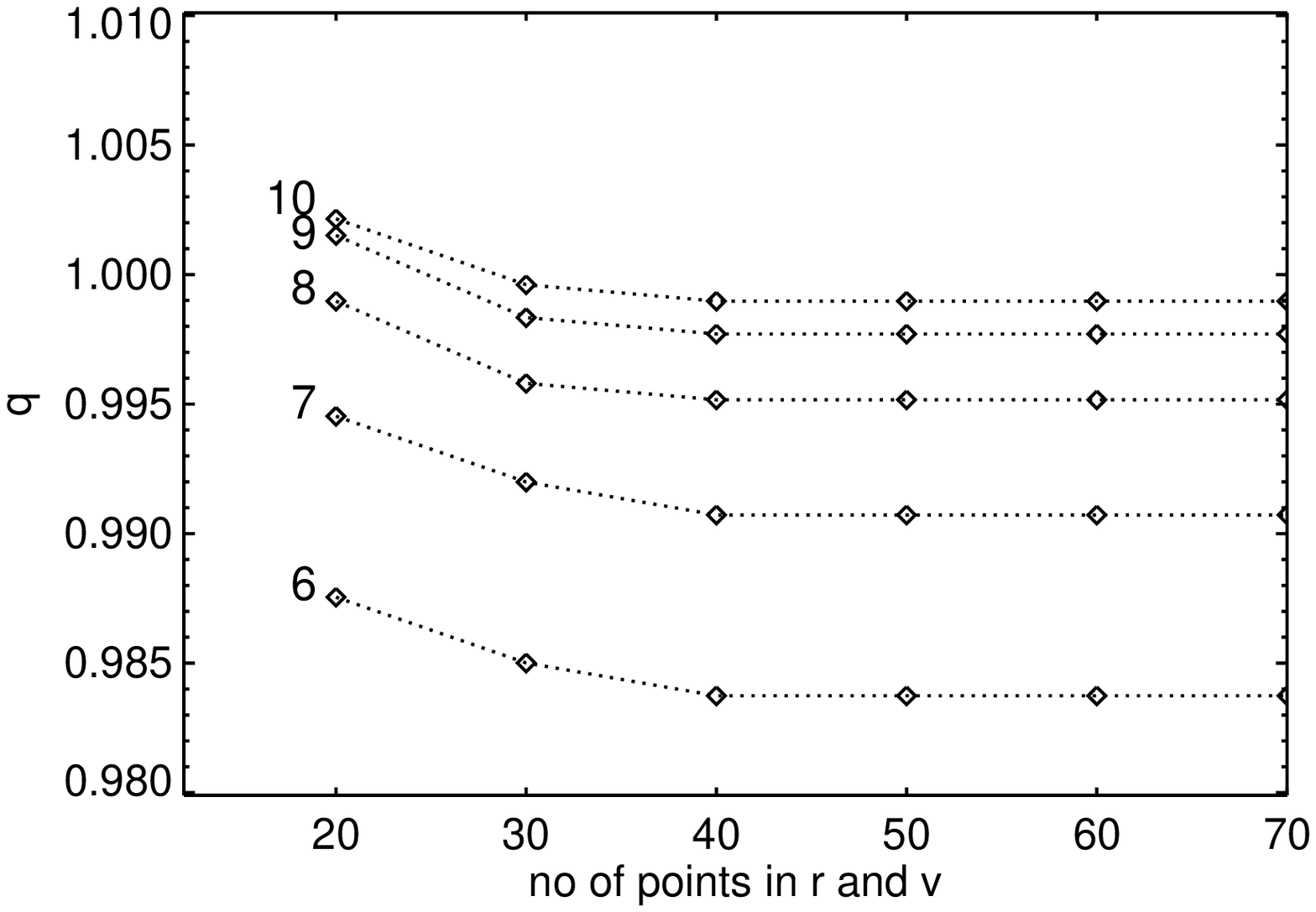}
\special{voffset=0 hoffset=235 hscale=40 vscale=50 psfile=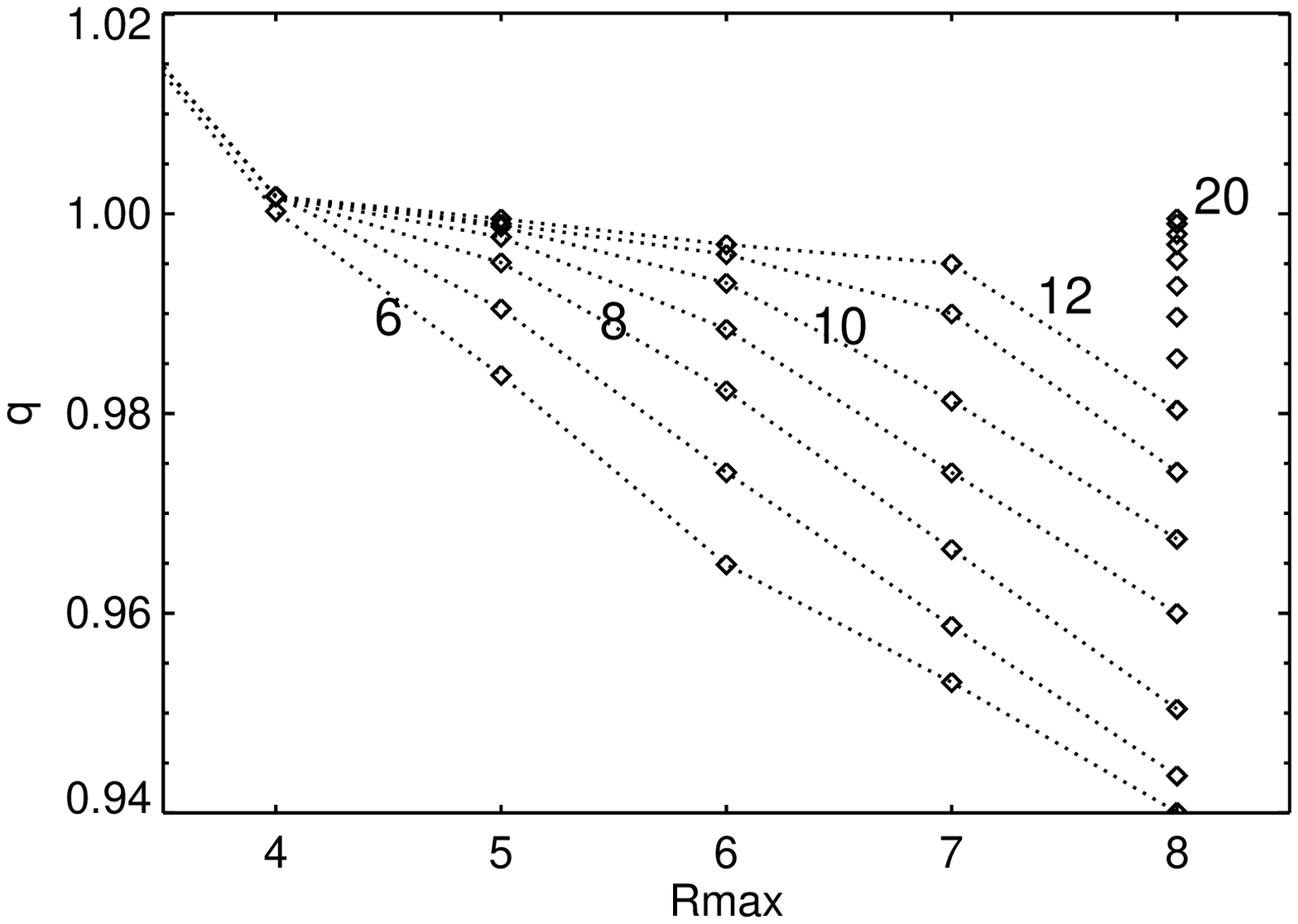}
\Caption
{The self-gravity parameter, $q$ as a function of the grid sampling, and the
 disk truncation.  Data  are shown for bases of  various sizes as  labelled.
 On the right,  $q$ is shown as  a function of  the truncation radius of the
 disk for  different sizes of basis as  labelled. More  functions are needed
 for convergence with  larger $R$ because the  inner, dominant,  part of the
 disk is less well sampled when the basis is stretched over a larger domain.
 As with all the  figures in this  section, the disk is Kalnajs' isochrone/9
 model.  The basis is Kalnajs' biorthonormal set with $k=7$.}
\endCaption
\endfigure 

\Fig{nrnv-rmaxplot} shows how the sampling affects the result when the
initial $R_0,  V_0$ grid has  the same number  of points  in each direction.
The   different lines  in  this  and   subsequent figures  indicate  how $q$
converges  for different  sizes  of basis.  The   figure shows that for this
disk, a sampling with a $40 \times 40$ grid reduces the sampling errors well
below those from the basis truncation.  The right-hand  figure shows how the
truncation of the disk influences convergence for  different sizes of basis.
If the instability  is localised to  the inner parts  of the disk, including
more of the outer regions  should not affect the  result. The behaviour seen
here is due to stretching of  the basis when $R_{\rm  max}$ is increased, so
more functions  are required  to   sample the  inner  regions  to  the  same
resolution.     If, however, enough functions are    used, no change is seen
beyond about $R_{\rm max} = 4.5$.

\midfigure{k-lplot}
\vskip 6.5cm
\special{voffset=0 hoffset=20 hscale=40 vscale=50 psfile=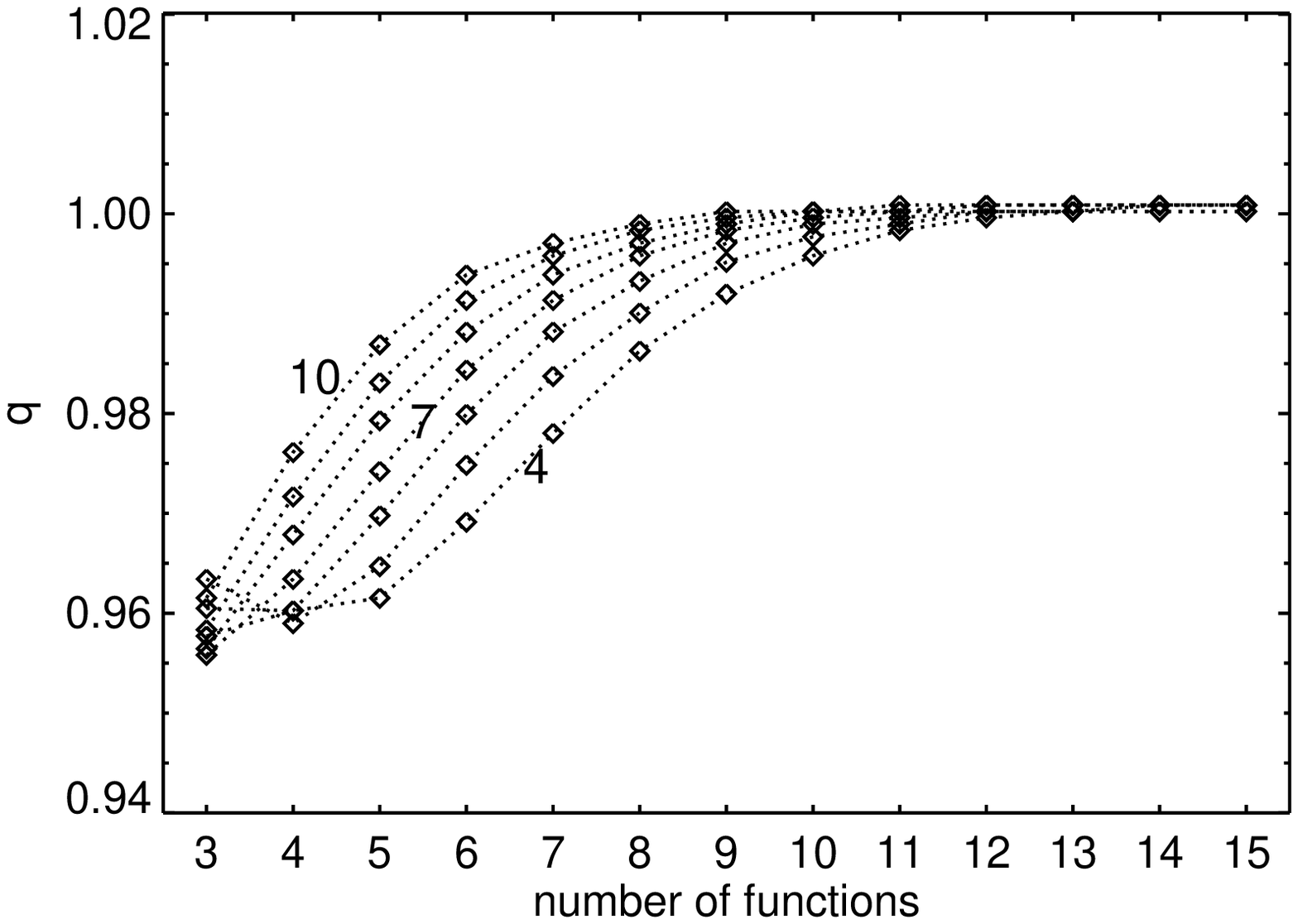}
\special{voffset=0 hoffset=235 hscale=40 vscale=50 psfile=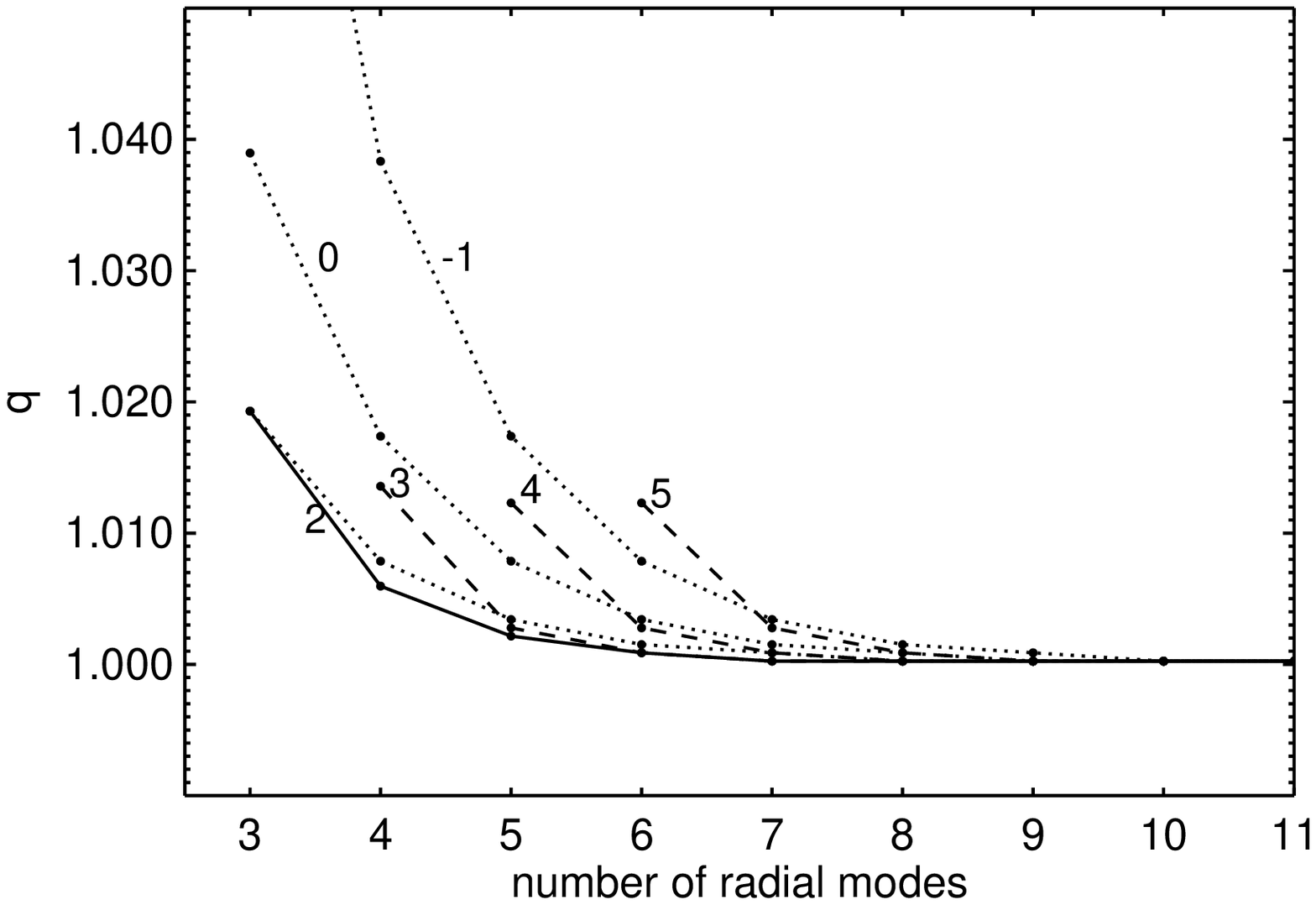}
\Caption
{The dependence of  the  self-gravity parameter,  $q$, upon the  order of the
basis, the number of  functions used in the calculation,  and the number  of
radial harmonics.  The  left-hand  figure shows $q$ against  the  basis size for
different bases  from Kalnajs'  set,   labelled  by the  corresponding   $k$
parameter. On  the right, 12 functions were  used, with the number of radial
harmonics on the abscissa taken on either side of a central value as labelled on
the   lines. Centering  the harmonics   on   $2$ (when  looking  for  bi-symmetric
instabilities) shows appreciably better convergence than taking equal numbers
of positive and negative harmonics (centre 0).}
\endCaption
\endfigure

As remarked by Earn   and Sellwood (1995)\cite{earn} choosing an  appropriate basis
for a given instability has   a significant effect   on the number of  basis
functions necessary    to resolve the mode.   This may be  seen in
\Fig{k-lplot} for Kalnajs' set of biorthonormal  bases with $  4 \leq k \leq
10$. Ten functions with the $k=4$ basis are required to  give the same error
in $q$ as achieved  with only six functions  in the $k=10$  basis.  Kalnajs'
remarks that  some economies may  be  made by not using  the  same number of
positive and negative radial Fourier modes but by centering the range on $m$,
the order of the instability. This is  illustrated in the right-hand figure,
where the  abscissa gives  the  number of  radial  modes on  each side  of a
central  value  as  labelled.  The computational   load   is almost directly
proportional to the number of radial modes to be used so this effect is well
worth exploiting.  Developing  specific basis for  different
thin disks also speeds up the calculation,  by reducing the number of 
functions required for convergence but it  is not necessary. Any
reasonable basis will get there in the end and the feasibility of the
resulting computations  considerably enhances the generality of this approach.

\midfigure{eigenvector-plot}
\vskip 6.cm
\special{voffset=0 hoffset=20 hscale=40 vscale=35 psfile=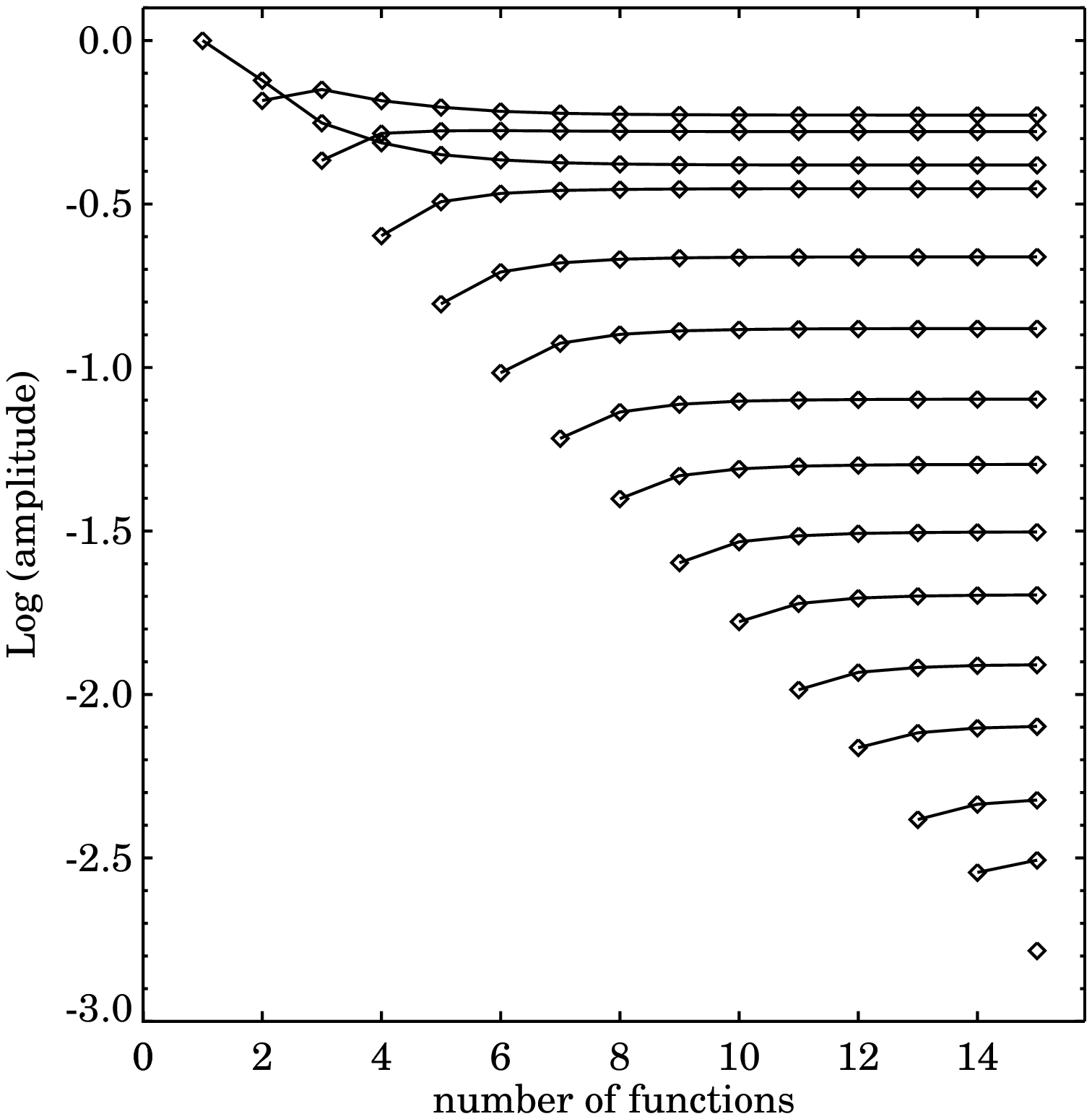}
\special{voffset=0 hoffset=235 hscale=40 vscale=35 psfile=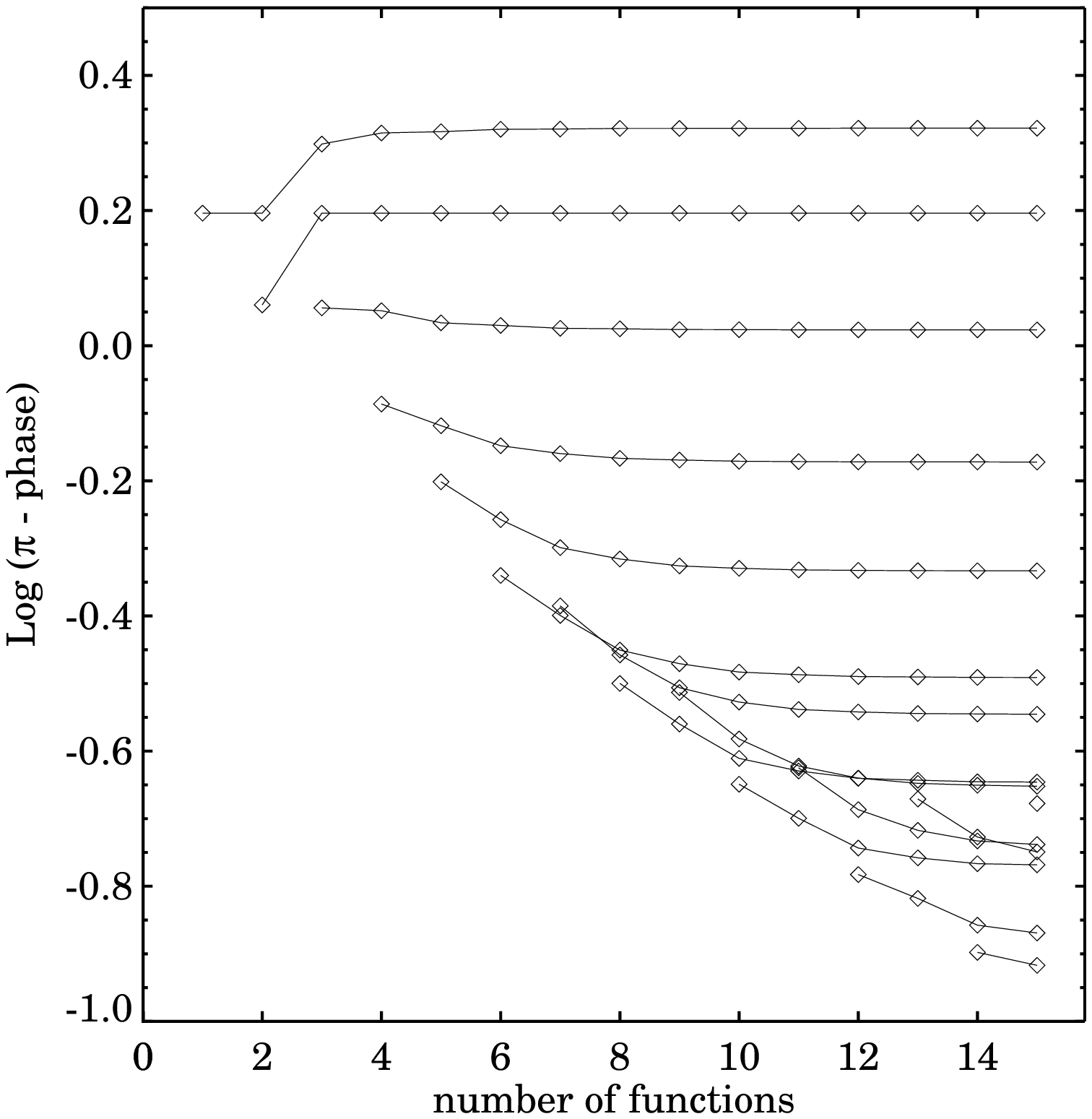}
\Caption
{The amplitude (left  panel) and the phase (right  panel)  of components of
the critical eigenvectors  as a function  of the number  of 
basis functions.
 The rapid convergence of the  shape of the  mode appears clearly.}
\endCaption
\endfigure

\section{ Application: The isochrone and Kuzmin-Toomre disks}

The versatility of the algorithm for linear stability described above is now
illustrated  while recovering growth rates  and  pattern speeds of  previously
studied disk with   old and new  distribution functions,   finding the modal
response -- in position space   and in action space  -- of these disks
both for bi-symmetric and lopsided modes.

\figure{dfdlb}
\vskip 7cm
\special{voffset=-10 hoffset=0 hscale=40 vscale=40 psfile=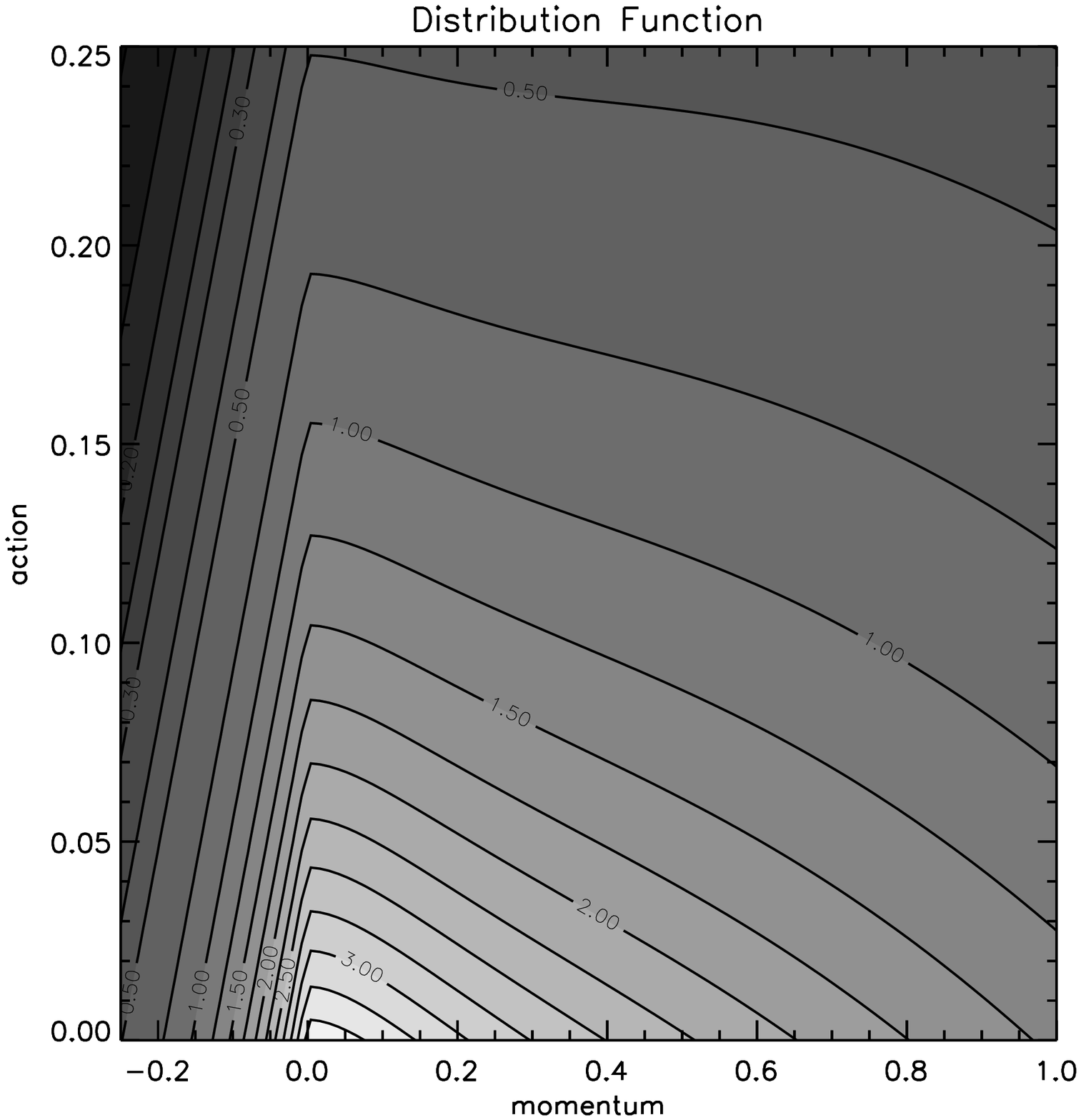}
\special{voffset=-10 hoffset=220 hscale=40 vscale=40 psfile=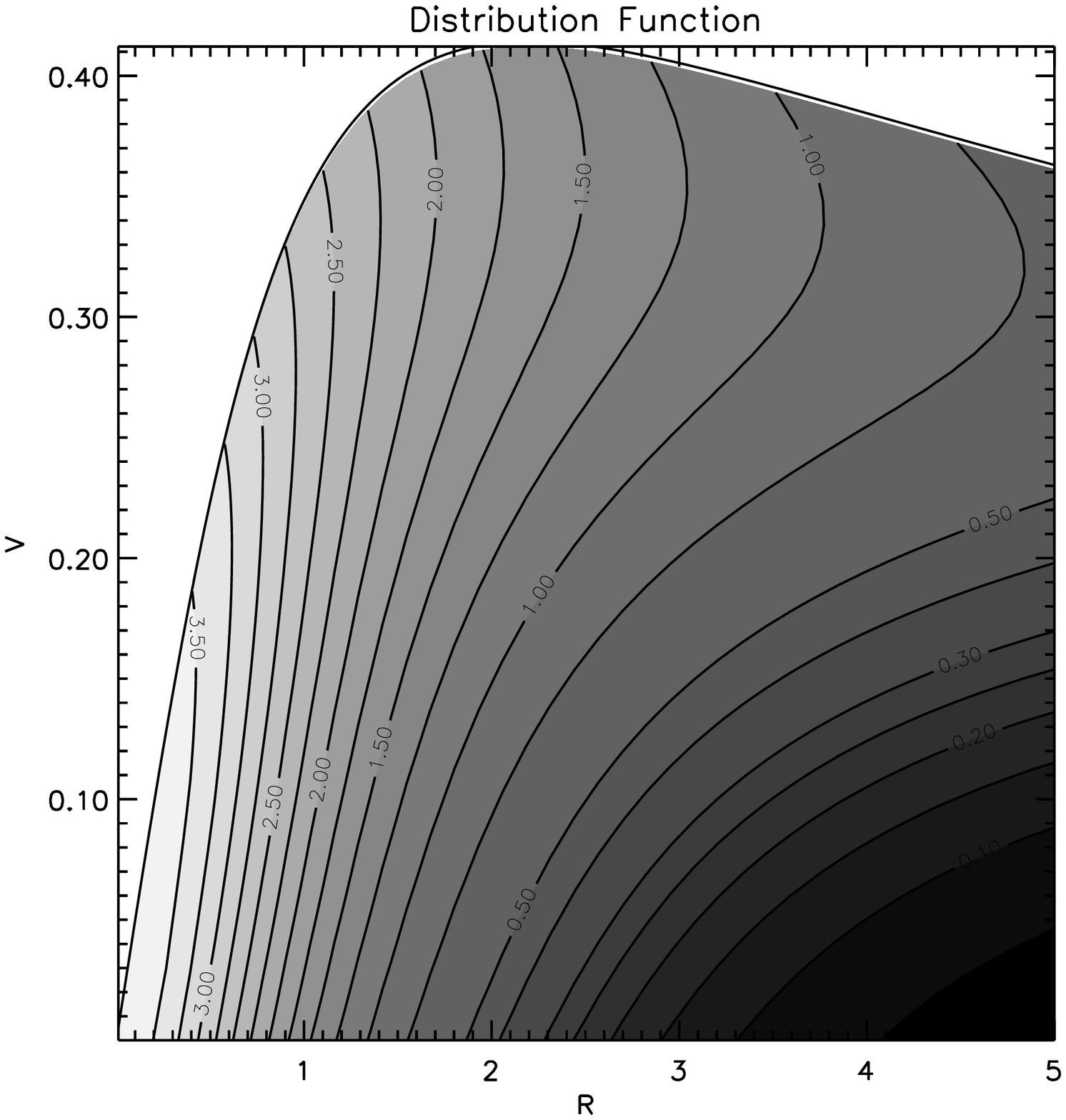}
\Caption
{left panel: isocontours  of the $m_{P}=4$  distribution  function in action
space.  The sharp break at  zero momentum is  an artifact of Kalnajs's trick
which was used here to incorporate counter rotating stars; right panel: same
isocontours   for  the prograde stars  only  in   $(R_{0},V_{0})$ -- apocentre,
velocity at apocentre -- space.  These variables  are those used throughout the
calculation to label the  orbits.  Note  the  envelope corresponding  to the
rotation curve  of that galaxy.  It appears  clearly in this parametrisation
that in this galaxy most orbits are of low ellipticity.  }
\endCaption
\endfigure 

\subsection{The Equilibria}

 \table{growthRateIso}
\caption{$m=2$ growth rates and pattern speed of the  isochrone/$m_K$ model.}
\intablelist{glop.}
\singlespaced
\ruledtable
\multispan3\hfil The  isochrone/$m_K$ model: bi-symmetric mode \hfil\CR
 Model \dbl $ m \Omega_p$ | $\eta$ \cr
 ~~~6 \dbl 0.34  | 0.075   \cr
 ~~~7 \dbl 0.37  | 0.085   \cr
 ~~~8 \dbl 0.43  | 0.125   \cr
 ~~~9 \dbl 0.47  | 0.145   \cr
~~~10 \dbl 0.50  | 0.170   \cr 
~~~11 \dbl 0.53  | 0.195  \cr 
~~~12 \dbl 0.59  | 0.210
\endruledtable
\endtable

\figure{qetaomega}
\vskip 14cm
\special{voffset=200 hoffset=0 hscale=50 vscale=50 psfile=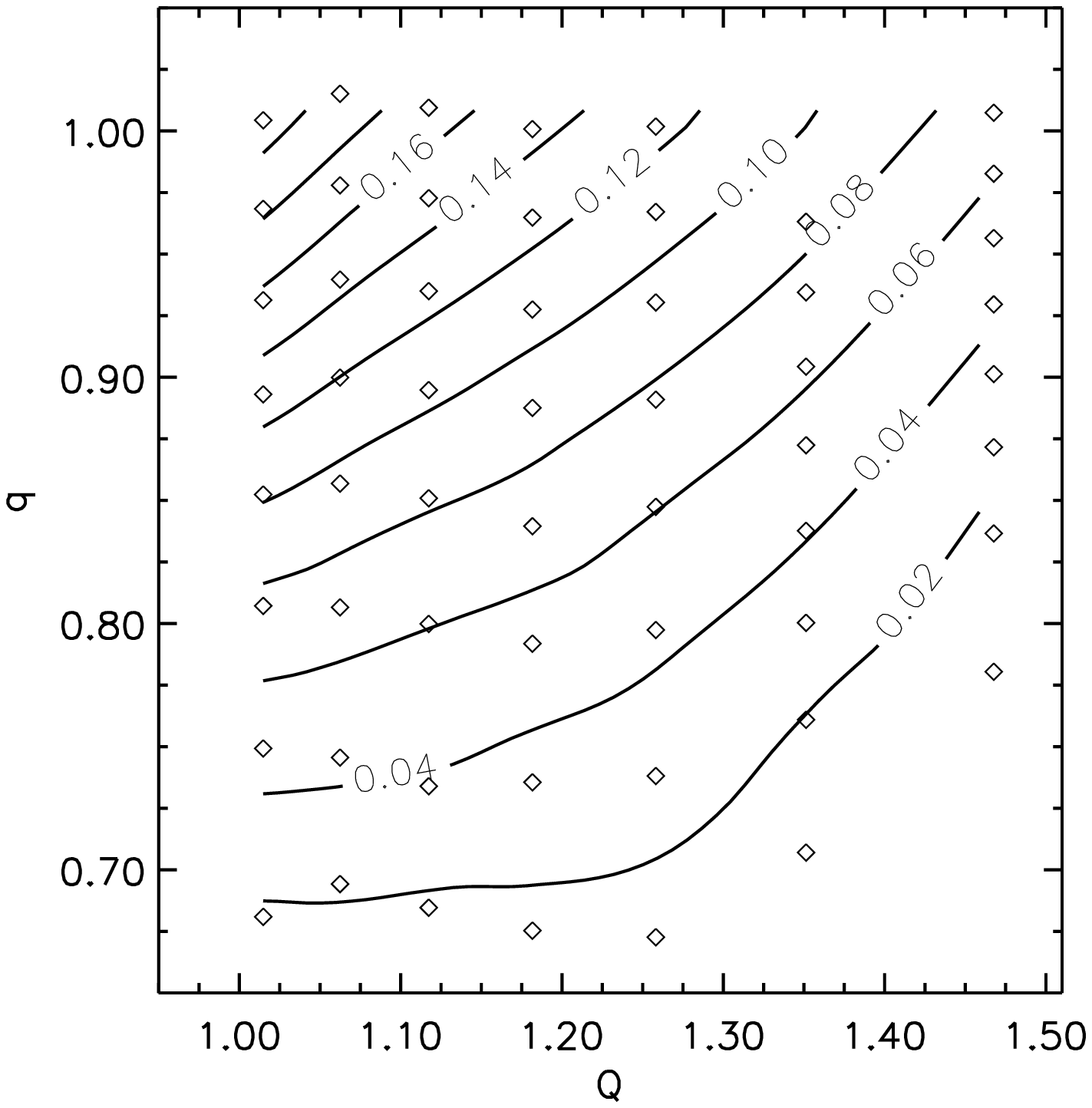}
\special{voffset=200 hoffset= 200 hscale=50 vscale=50 psfile=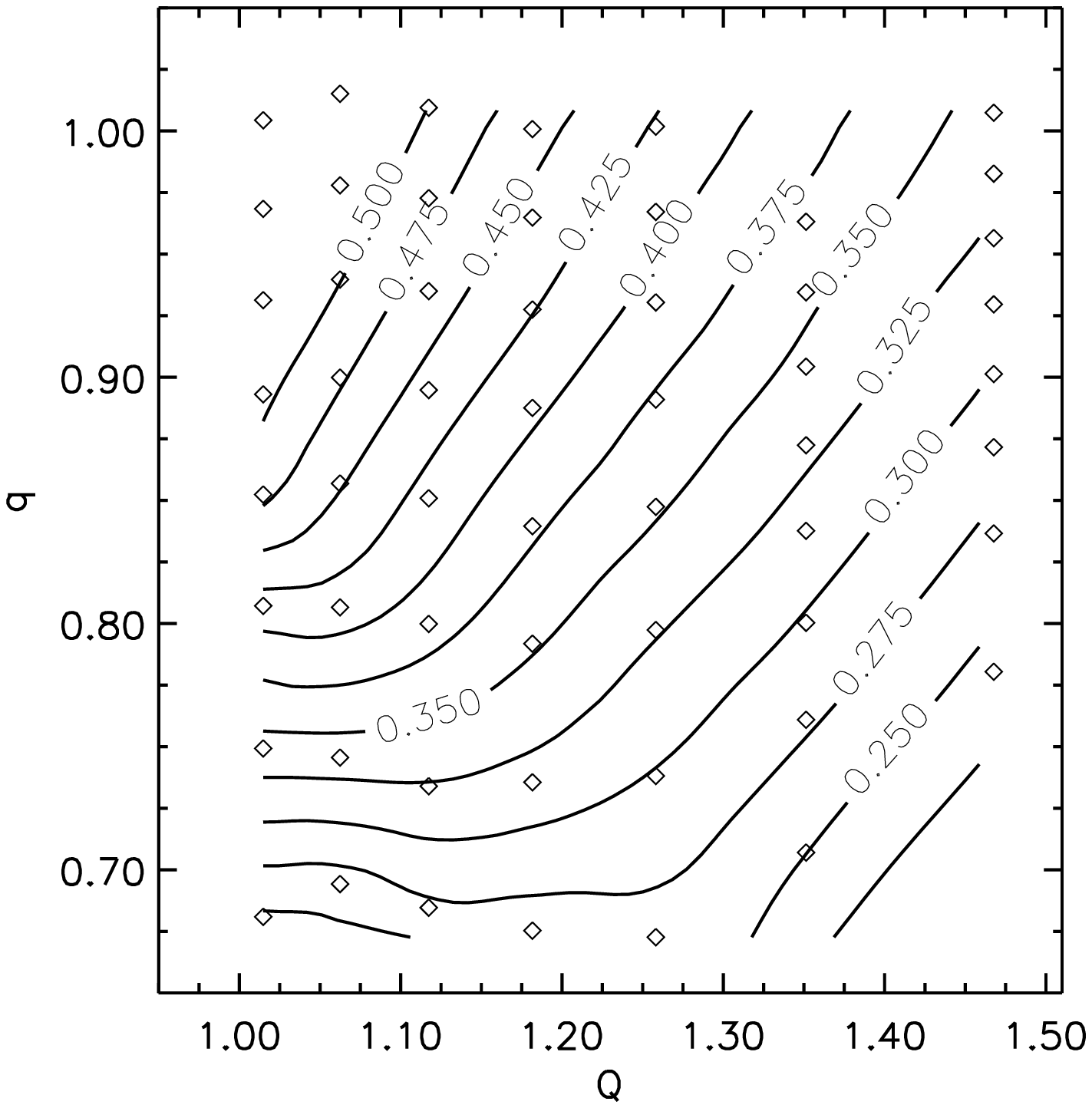}
\special{voffset=-10 hoffset=0 hscale=50 vscale=50 psfile=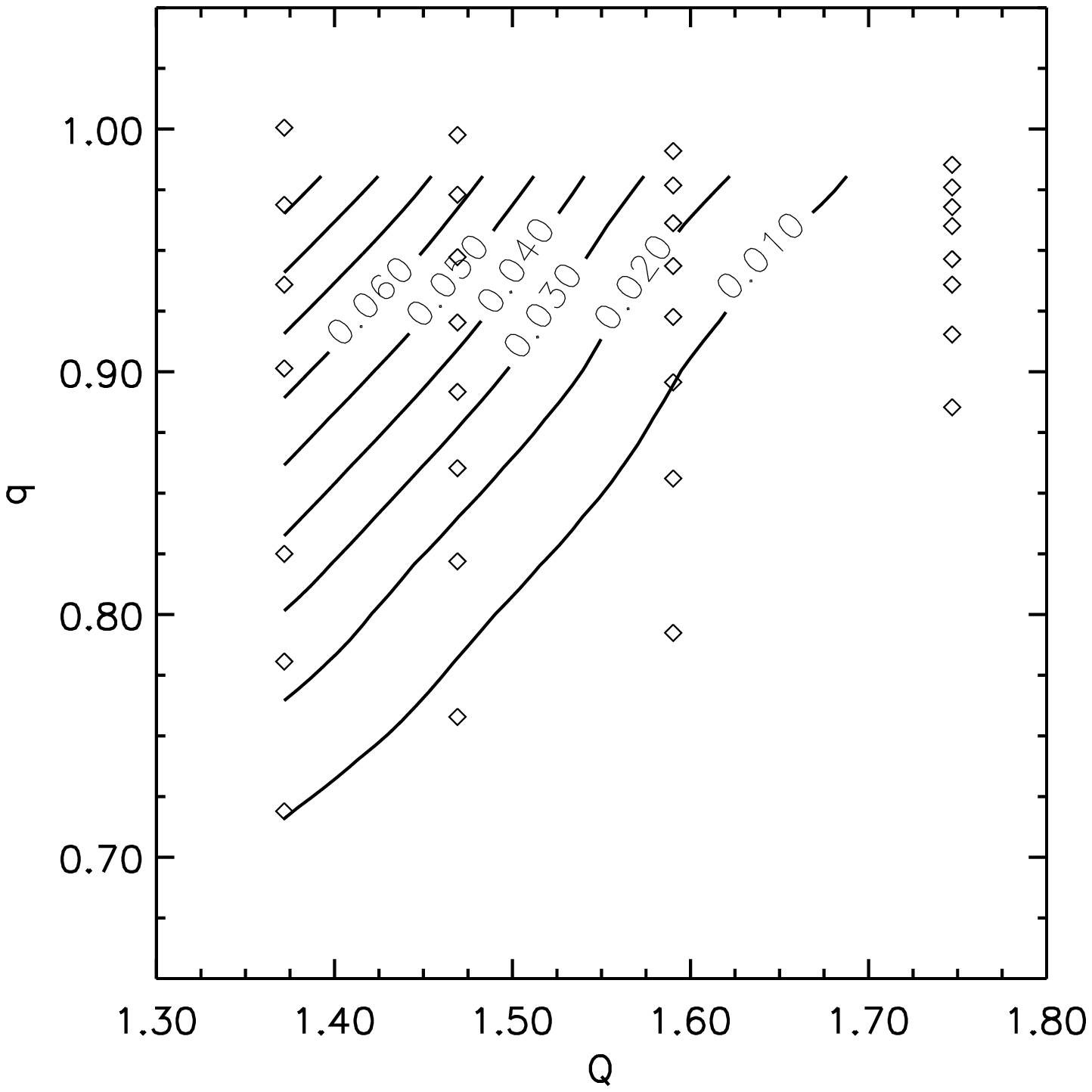}
\special{voffset=-10 hoffset= 200 hscale=50 vscale=50 psfile=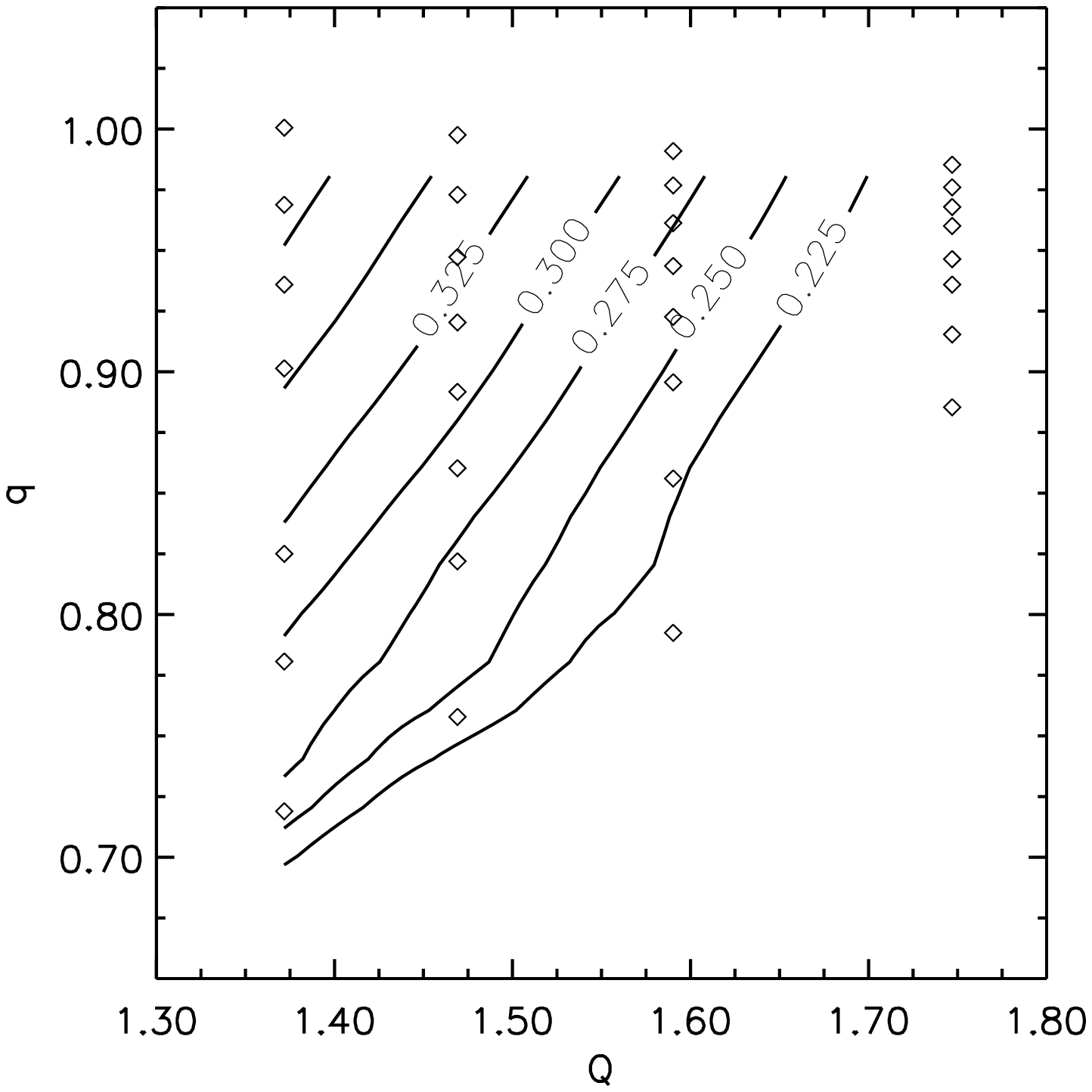}
\Caption
{the variation of the  $m=2$ growth rate (top  left panel)  and the pattern  speed
(top right panel) of  an isochrone/$m_{K}$ family as a  function of the self
gravity parameter $q$ and  the Toomre number $Q$.   The bottom  panels show
the same results for the isochrone/$m_{P}$  family.  The precision
in the growth rates --  at fixed truncation in  the basis -- drops for lower
values of $Q$ since these modes are more centrally concentrated. }
\endCaption
\endfigure 

\figure{qQfit}
\vskip 7cm
\special{hoffset=-40 voffset=-90 hscale=50 vscale=50 psfile=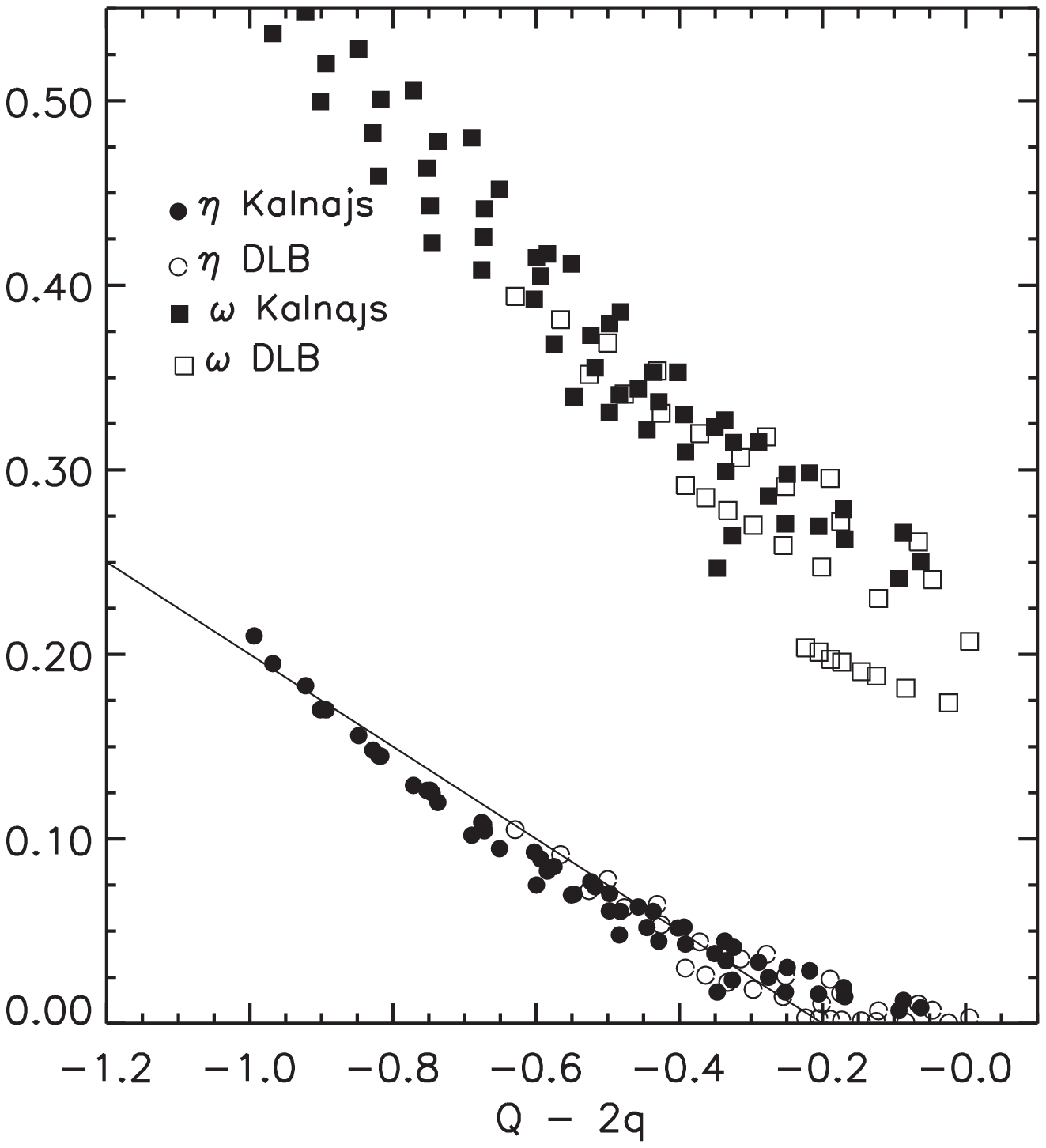}
\special{hoffset=200 voffset=-5  hscale=50 vscale=50 psfile=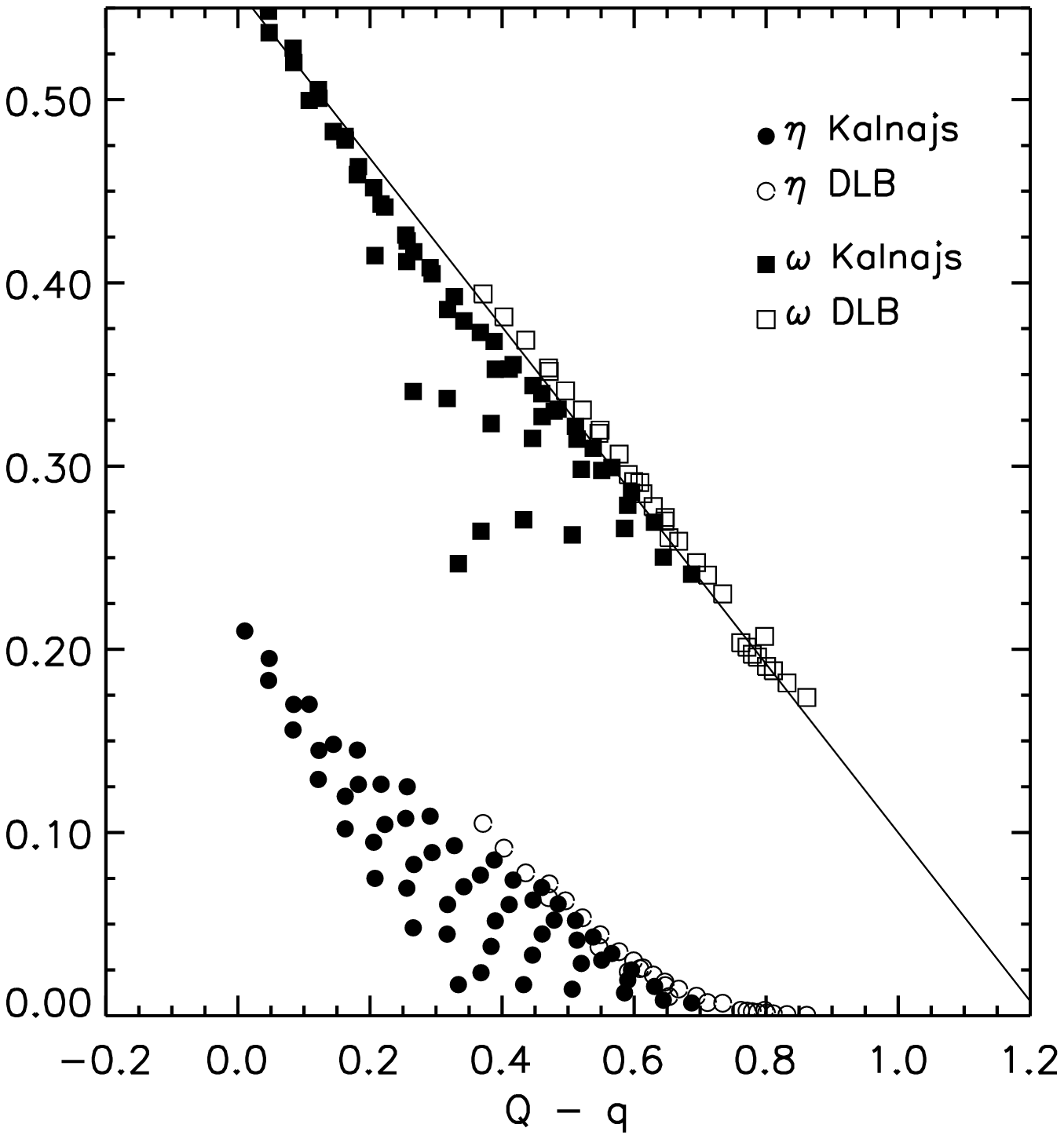}
\Caption
{fits to the evolution of the $m=2$ growth rate and the  pattern speed of an
isochrone/$m_{K}$ (marked Kalnajs) and an isochrone/$m_P$ (maked DLB) family
as a function of a linear combination of the self  gravity parameter $q$ and
the Toomre  number $Q$.  The left panel  minimises  the dispersion in $\eta$
whereas  the right panel  that in $\omega$.   The best fits are obtained for
different linear combinations of  $Q$ and $q$. The solid  line for $\eta$ is
$\eta = -0.05 - 0.25 (Q - 2q)$ and that for  $\omega$: $\omega = 0.56 - 0.46
(Q-q)$.}
\endCaption
\endfigure

\table{growthRateDLB}
\caption{ $m=2$ growth rates and pattern speed of the isochrone/$m_{P}$ model.}
\intablelist{glop.}
\singlespaced
\ruledtable
\multispan3\hfil The isochrone/$m_P$ model: bi-symmetric mode \hfil\CR
Model \dbl $ m \Omega_p$ | $\eta$ \cr
 ~~~3 \dbl 0.21 | 0.003   \cr
 ~~~4 \dbl 0.29 | 0.032   \cr
 ~~~5 \dbl 0.35 | 0.072   \cr
 ~~~6 \dbl 0.40 | 0.105  
 \endruledtable
\endtable

\table{growthRateHunter}
\caption{$m=2$ growth rates and pattern speed of the  Toomre/$m_{M}$ model.}
\intablelist{glop.}
\singlespaced
\ruledtable
\multispan3\hfil The Toomre/$m_{M}$ model: bi-symmetric mode \hfil\CR
Model \dbl $m \Omega_p$ | $\eta$ \cr
 ~~~2 \dbl 0.598  | 0.204   \cr  
 ~~~3 \dbl 0.714  | 0.294   \cr  
 ~~~4 \dbl 0.810  | 0.371   \cr  
 ~~~5 \dbl 0.916  | 0.445        
 \endruledtable
\endtable

Three  families of disks are  studied here:  two corresponding to equilibria
for  the isochrone disk (1961)\cite{Henon},   and one for the  Kuzmin-Toomre
(1956)\cite{Kuzmin}  mass model.    The  isochrone disk  is defined   by its
potential, $ \psi =GM/(b+{r_b})  \quad {\rm where   } \quad {r_b}^2 =  R^2 +
b^2$.  The  corresponding surface  density is  $\Sigma ={{Mb}  }\left\{ {\ln
[(R+{r_b})/b]-R/{r_b}}
\right\}$ $/({2\pi R^3})$.
The Kuzmin-Toomre potential  is defined by 
$ \psi =GM/{r_b}$ and 
the corresponding   surface density is   $\Sigma ={{Mb}  / {2\pi r_b^3}}  $.
Miyamoto (1974)\cite{Miyamoto}, followed by Kalnajs  (1976)\cite{kalnajs76},
Athanasoula \&   Sellwood  (1985)\cite{atha85}, and  Pichon  \&  Lynden Bell
(1996)\cite{Pichon2} chose specific forms of distribution functions assuming
simple algebraic Ansatz   for  its  expression in   $(\varepsilon,h)$  space.
Historically,   these  models were  first   used  to  construct  families of
maximally  rotating   disks   ({\it e.g.}  the    Miyamoto/$m_M$   disk
models\cite{Hunter} ) while     for  others, counter-rotating   stars   were
re-introduced  by simple, though  somewhat  arbitrary,  tricks  (Kalnajs
(1978)\cite{kalnajs78} described by Earn \& Sellwood(1995)\cite{earn}).  In this paper,
The Kalnajs isochrone/$m_{K}$  models  are implemented  in order to  recover
Kalnajs     (1978)\cite{kalnajs78}   first   linear stability   results   on
differentially rotating disks.  To demonstrate the flexibility of the method
the truncated  version   of Hunter's Toomre-Kuzmin/$m_{M}$  models  are also
constructed, and the stability of the equilibria given by Pichon \& Lynden
Bell (1996) are analysed.

Kalnajs's distribution family reads  (in units of $b$ and $G=1$)
$$ f_K(\varepsilon,h)= 2^{2 m-1} (  \sqrt{-2 \varepsilon } h)^{-m}  \pi^{-1}
(-\varepsilon)^{m-1} g( \sqrt{-2 \varepsilon } h) \, ,\EQN kaldf $$ where $$
g(x)= x {\partial (x^m \tau_m(x))\over  \partial x} +\int_0^1 (t \, x)^m \tau_m(t
\, x) {\rm P}''_{ m-1}(t)\, \d t -\half {( m-1) } m x^m \tau_m(x) \, , \EQN $$ and $$
\tau_m(x) =\left.  {
\log(r+\sqrt{1+r^2})- r/\sqrt{1+r^2} \over 2 \pi r^3 
(-1-\sqrt{1+r^2})^{-m}} \right|_{ r= 2x /(x^2-1)}.  \EQN $$ 
Here ${\rm P}_{m}$ 
stands here for the Legendre polynomial of order $m$.  Pichon and Lynden-Bell's
distribution family for the isochrone disk is given by 
$$f_P(\varepsilon,h)=\left.{\left( {{{-\varepsilon } {}}} 
\right)^{m+1/2}
\over  {4\pi  ^2(m+1/2)!!  }}{{{h} \over  {\sqrt2}}}\left(  {{{\partial ^{}}
\over {\partial ^{}s}}} \right)^{m+2}[(s^2-1)^m L(s)]\right|_{s = 1 +h^2/2}
\, , \EQN PLBdf$$
with  $ L(s)=  \log(\sqrt{s^2-1}+s)   +\sqrt{s^2-1}/s $.  Its   contours for
$m_{P}=4$ are illustrated in \Fig{dfdlb} while the $Q$ profiles are given by
Pichon \&  Lynden Bell (1996)\cite{Pichon2} (Fig~PLB~5).  Finally Miyamoto's
distribution is
$$f_M(\varepsilon,h)={ \left(  2\,m+3 \right) \over 
{2\,{{\pi }^2}}}{{{{\left( -\varepsilon \right) }^{2 + 2\,m}}\, \,
{ {}_2 F_1}(-m,-2 - 2\,m,{1\over 2},{{-{h^2}}\over 
{2\,\varepsilon}})}}  \, ,
\EQN hunterdf$$ where ${ {}_2 F_1}$ is  a terminating  Hypergeometric 
function of the second kind.

\figure{densityResponse}
\vskip 7cm
\special{voffset=-20 hoffset=0 hscale=40 vscale=40 psfile=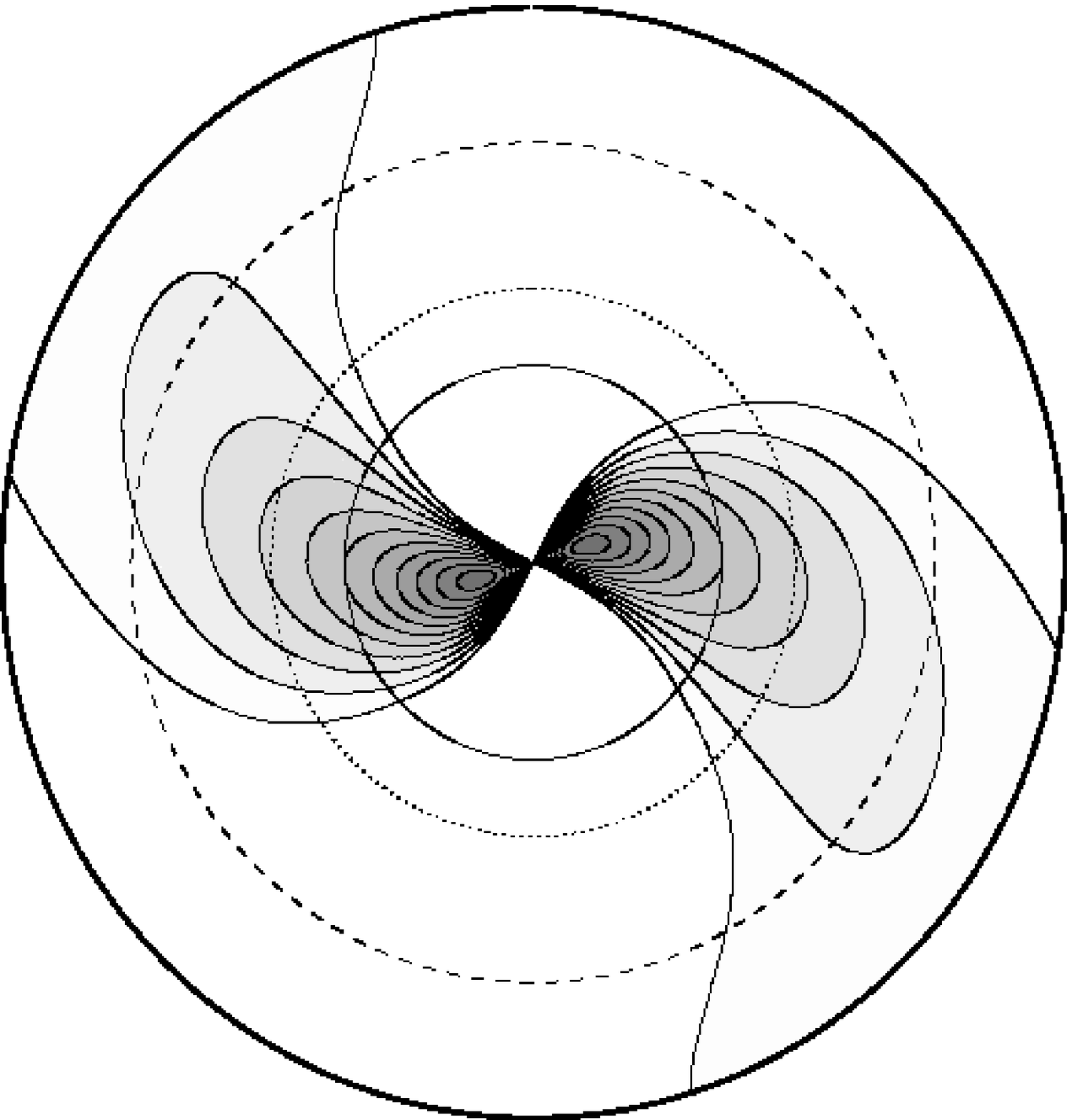}
\special{voffset=-20 hoffset= 200 hscale=40 vscale=40 psfile=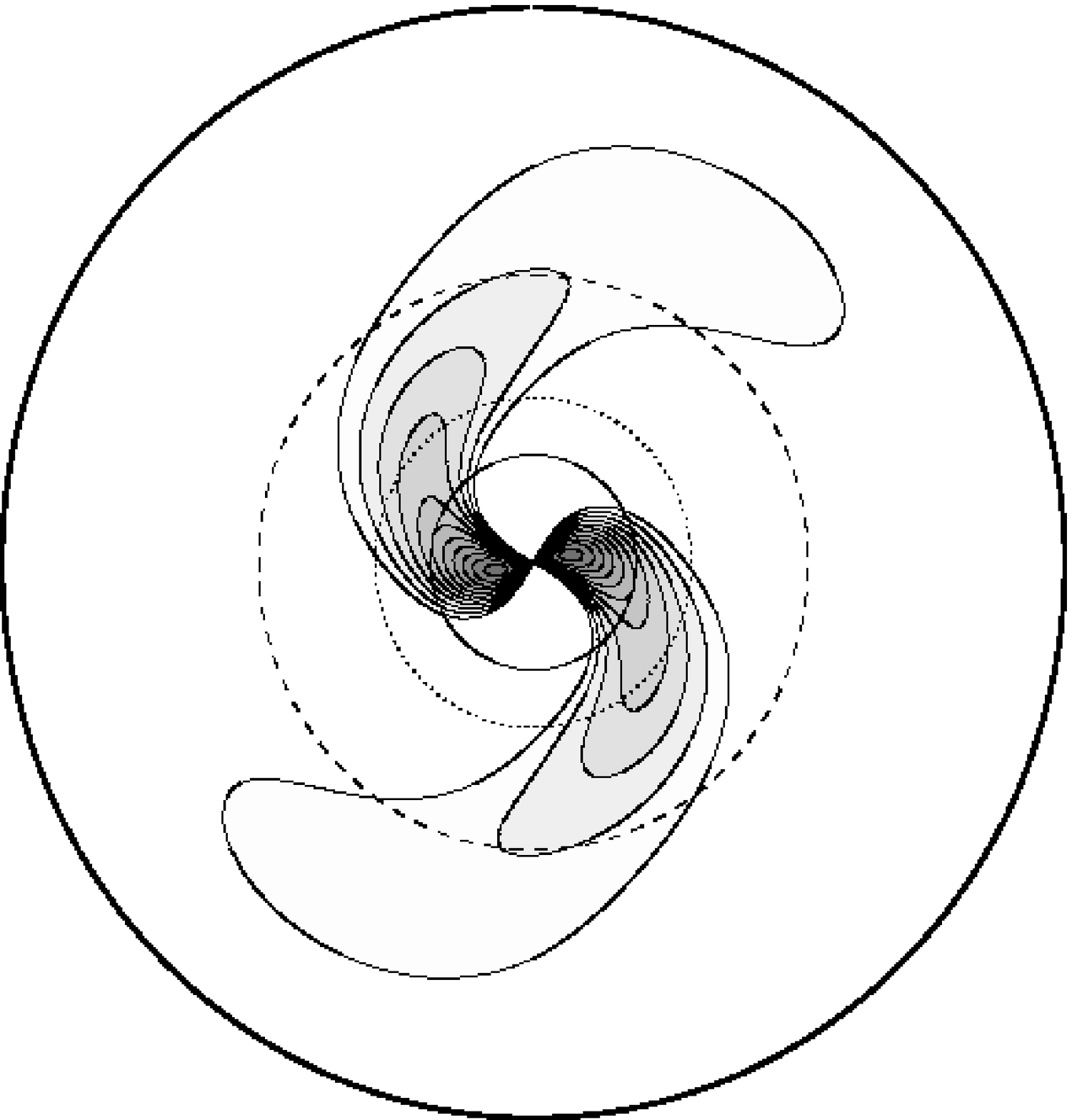}
\Caption
{the density response corresponding to the fastest $m=2$ growing mode of the
isochrone$/7$ and isochrone$/11$ model.  Note  that  the colder $/11$  model
yields    a  more tightly    wound    spiral  as  shown   quantitatively  in
\Fig{WaveShape}. This is expected since in the locally marginally radially
unstable r\'egime, the disk   response should asymptotically match  that  of
unstable  rings.  Its  spiral response is   also more centrally concentrated
than  that of its  hotter  counterpart.   The  solid  circle  corresponds  to
$R_{1/2}$, the radius at which the  wave has damped by a  factor of two; the
dotted   circle to Corotation   resonance; the  dashed circle  to  the outer
Lindblad resonance and the outer circle to $R_{max}=5$. }
\endCaption
\endfigure

\figure{densityResponse2}
\vskip 7cm
\special{voffset=-20 hoffset=80 hscale=40 vscale=40 psfile=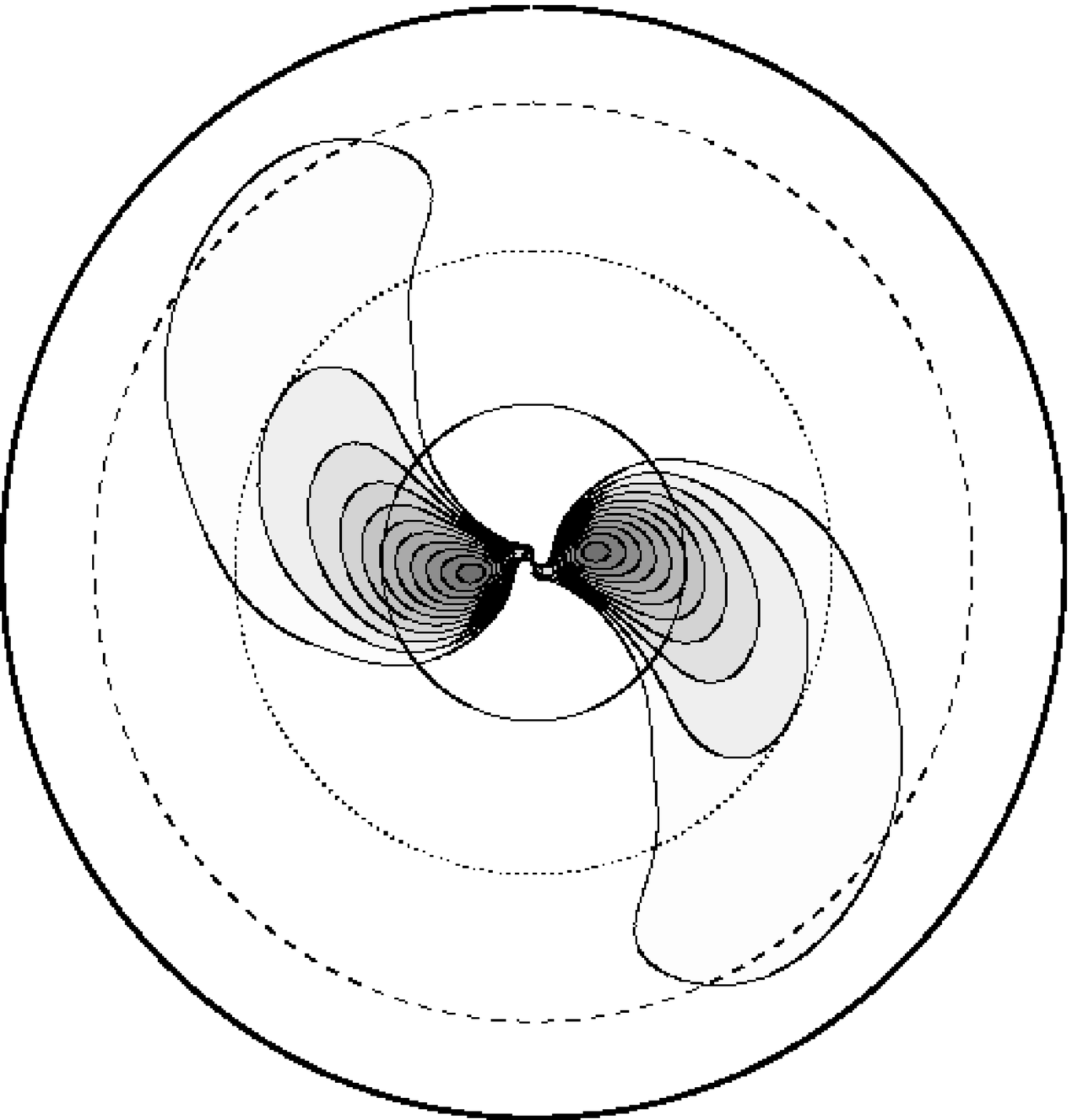}
\Caption
{ same as \Fig{densityResponse} for a
Myamoto-Hunter$/3$   model. Note that the response is much more centrally
concentrated since the outer circle is $R_{max}=3$. This is expected
since the Kuzmin-Toomre potential is more compact than the isochrone.}
\endCaption
\endfigure

\subsection{ Characteristics of the  linear wave}

\figure{WaveShape}
\vskip 7cm
\special{voffset=-80 hoffset=80 hscale=40 vscale=40 psfile=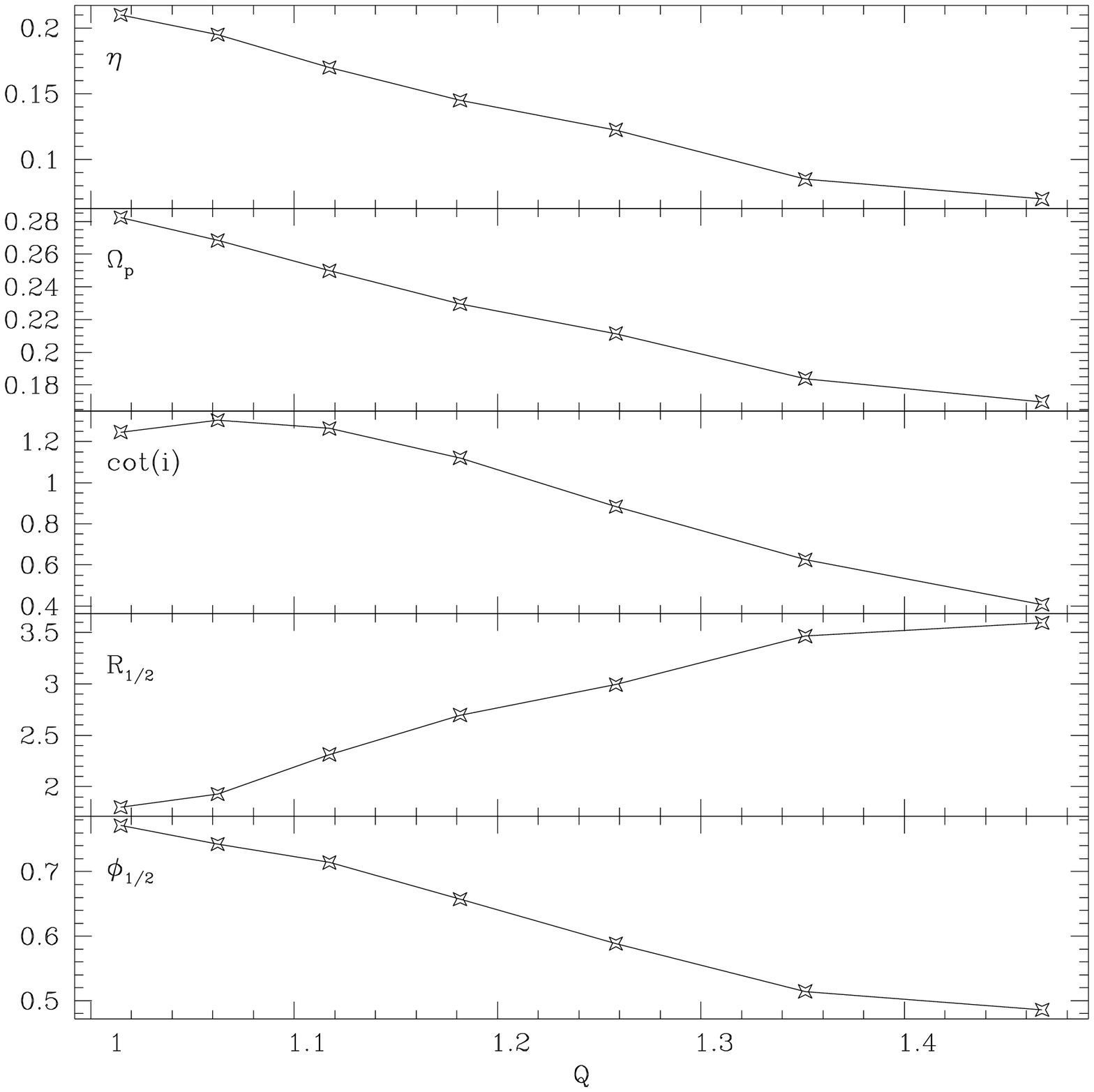}
\Caption
{the  evolution  of $\eta$  ,$\Omega_{p}$,  $   \cot(i)$, $R_{\rm 1/2}$  and
$n_{\rm 1/2}$, as a function of the Toomre number $Q$ of the isochrone/$m_{K}$
disks. Note that the hotter disk have less wound spiral response. }
\endCaption
\endfigure

 \table{secondgrowthrateIso}
\caption{ second and third fastest $m=2$ growing modes of the isochrone/$m_K$ model.}
\intablelist{glop.}
\singlespaced
\ruledtable
\multispan3\hfil The  isochrone/$m_K$ model: bi-symmetric mode \hfil\CR
 Model  \dbl second mode |  third mode \cr  ~~~7 \dbl  0.29(4)  + 0.04(1)i |
 0.22(0) + 0.008(8)i \cr ~~~10 \dbl 0.39(7) + 0.10(5)i | 0.27(3) + 0.039(5)i
 \cr ~~~11  \dbl  0.42(8) + 0.12(8)i |  0.34(7)  + 0.091(5)i \cr ~~~12  \dbl
 0.46(1) + 0.14(5)i | 0.26(4) + 0.051(2)i
\endruledtable
\endtable

\subsubsection{Bi-symmetric  $m=2$  modes}

The pitch angle of the spiral response is commonly defined as $
\cot(i)=\left\langle{\partial \theta / \partial \log(R)} \right\rangle_{\theta}$
where  $\theta=\theta(R)$ at the crest of  the spiral wave.  In practice, it
is best to calculate $ \tan(i)= \left\langle{\partial \log(R) / \partial
\theta }
\right\rangle_{R}$  since   $R=R(\theta)$  is a bijection   in  $[0,2 \pi/m[$.
Another  useful set of quantities are  defined by the radius, $R_{\rm 1/2}$,
(resp.  the  angle $\theta_{\rm  1/2}$) at  which  the spiral  response  has
decreased      to half it      maximum   amplitude.    The   winding  number
$n_{1/2}=\theta_{\rm 1/2}/(2\pi)$ yields  a  measure of  the winding of  the
wave: the larger $n_{\rm 1/2}$ the more wound it is.  Comparing the position
of the resonances   and $R_{\rm max}$ to   $R_{\rm 1/2}$  provides  means to
assess whether the   truncation of the disk   is  likely to  have  generated
spurious   cavity waves.  Specifically, it  is   required that $R_{\rm 1/2}<
R_{\rm OLR}  < R_{\rm max}$ so  that the wave is   well damped by  the outer
Lindblad resonance before the disk  is truncated.  Numerical simulation have
suggested that $R_{\rm 1/2} \sim R_{\rm COR}$.  Both statements are verified
here  for instance in  \Fig{densityResponse} (resp.  \Fig{densityResponse2})
which gives the linear  response of a $m_{K}=7$  and a  $m_{K}=11$ isochrone
disk   (resp.      $m_{K}=5$ Kuzmin disk)   for    its   first growing mode.
These perturbations display the usual bar shaped central response with a
loosely wound spiral response further out.
\Fig{WaveShape} gives the evolution of  $  \cot(i)$, $R_{\rm 1/2}$,  $n_{\rm
1/2}$, $\Omega_{p}$, $\eta$, as  a function of  the Toomre number $Q$ of the
Kalnajs /$m_{K}$ disks.
\Tab{growthRateIso}-\Tab{growthRateHunter} give the growth rates and 
pattern speeds of the $/m_{K}$  $/m_{P}$ and  $/m_{M}$ families.  Note  that
for the new $/m_{P}$ family the  bar mode given in \Tab{growthRateDLB} again
have pattern speed well above the maximum of $\Omega -  \kappa/2 $ and hence
do not  display inner Lindblad resonance.  Note  also  that the more compact
Kuz'min  -Toomre  potential has larger   growth rates  and pattern speed  as
expected since the  dynamical time is shorter and  the self gravity enhanced
for those disks.  
\Tab{secondgrowthrateIso} gives the 
growth rates for the second and the third growth  rate of the $m_{K}$ disks.
Note that the growth rate of the second fastest growing mode of a $m_{K}=12$
model  was found by this  method to equal $\omega +  i\eta = 0.461 +0.145i$,
roughly  within the error bars  given by Earn \& Sellwood (1995)\cite{earn}.
Note  that the number  of radial nodes  for these slower modes increase with
the  rank  in stability  (as illustrated in  \Fig{densityResponse1-2}  for a
$m=1$ mode discussed in the next subsection) which  is consistent insofar as
winding decreases self gravity.  Note also the  small relative error between
the growth rates recovered for the  Myamoto/Hunter models and those given by
Hunter  (1993)\cite{Hunter}  for the   Toomre-Kuzmin disks.  Small  residual
discrepancies are expected  given that Hunter's  disks are  infinite whereas
those studied here are truncated at $R = 5$.
\Fig{qetaomega} gives for the $m_{K}$ and  $m_{P}$ families the evolution of
the first growing mode as a function of the Toomre $Q$ number (mass averaged
in the  inner region $R<2$)  of these disks and  ``the mass in the halo'' as
parametrised by $q$.  An {\sl new} attempted fit of a combination of $Q$ and
$q$ is also given in
\Fig{qQfit} for both distributions;  the relative dispersion illustrates the
well known  fact  that the   stability does  not  depend  only  on a  simple
combination of   these numbers.  These  curves   seem to  be  in qualitative
agreement  with  those  given by  Vauterin  \&  Dejonghe\cite{Dejonghe}  for
different disk models.  Here too, the pattern  speed at fixed growth rate is
a decreasing  function of  the  $Q$ number  ($\partial  \Omega_p /  \partial
Q_\eta < 0$).  The conjecture  of Athanasoula \& Sellwood\cite{atha85} of an
asymptotic value of $Q\approx 2$ for marginal stability  of { \sl fully self
gravitating} disk seems also consistent with these results.

\figure{orbitalResponse}
\vskip 7cm
\special{voffset=-10 hoffset=0 hscale=40 vscale=40  psfile=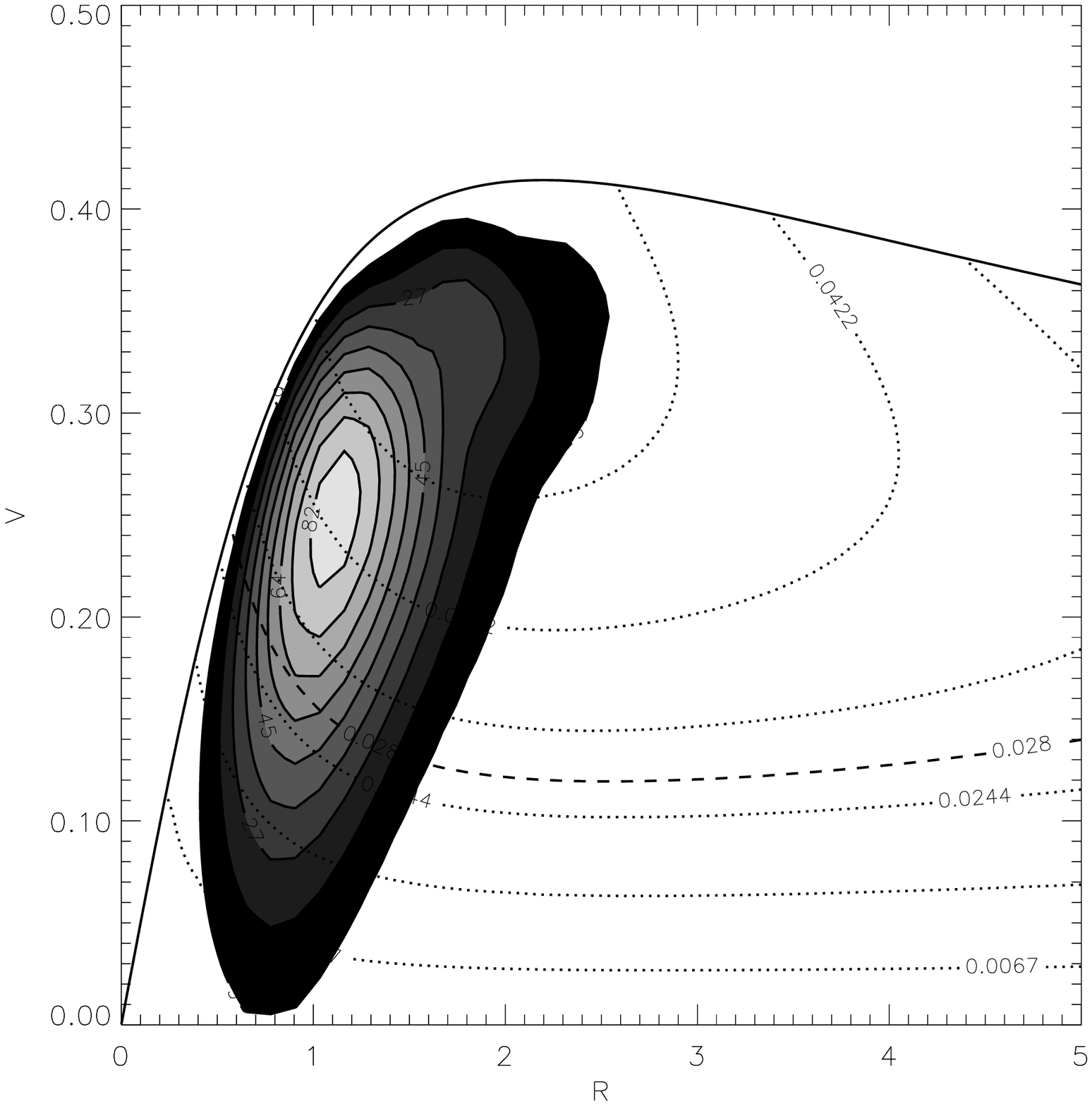}
\special{voffset=-10 hoffset= 210 hscale=40 vscale=40 psfile=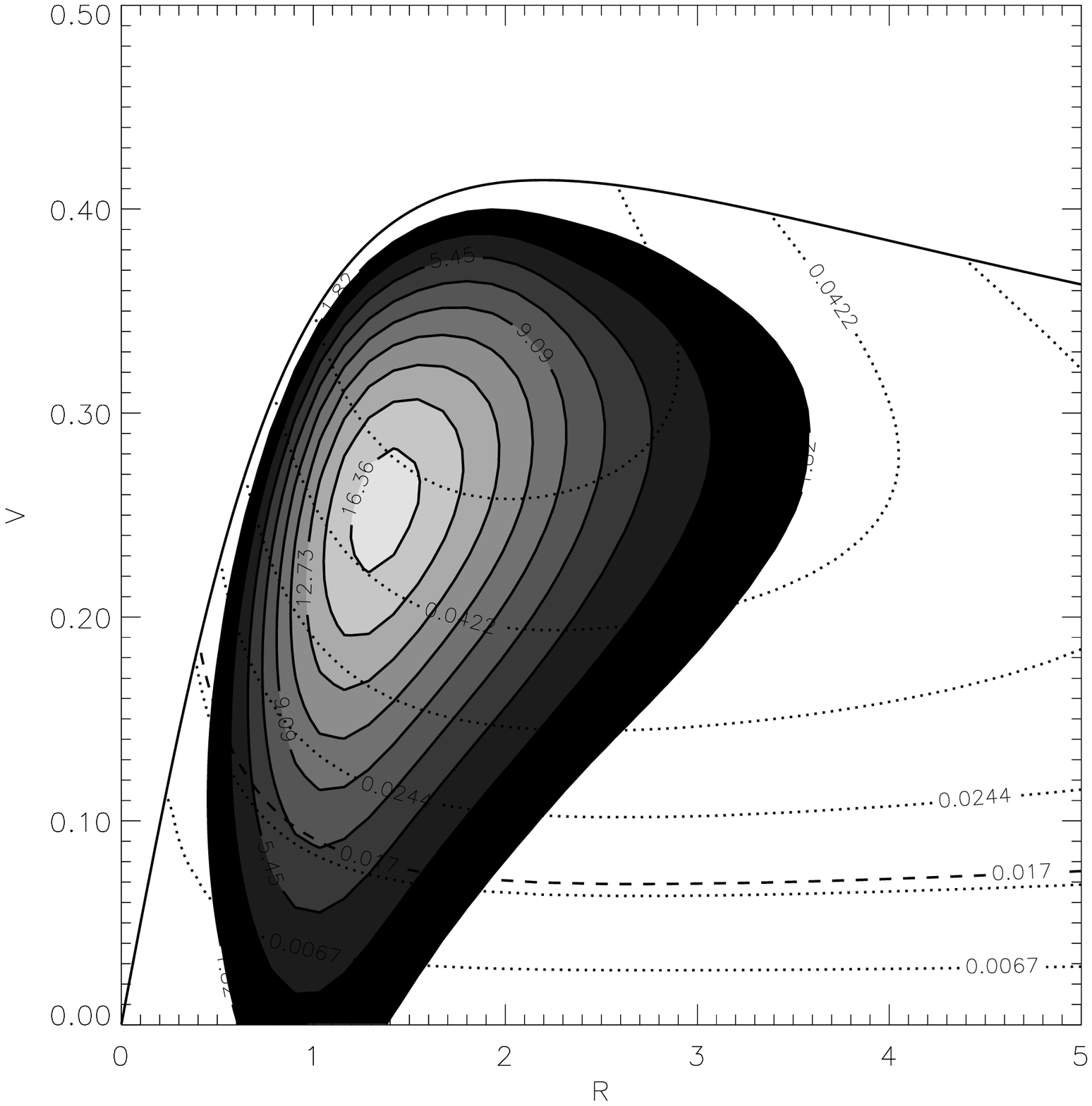}
\Caption
{the  orbital response in action  space corresponding to the fastest growing
mode described  in  \Fig{densityResponse}.  The  amplitude  of the left hand
side of \Eq{4-14} is plotted for the $\ell=-1$ (ILR) fastest growing mode of
an isochrone/12 model   (left  panel), and  an  isochrone/6 model.    (right
panel).  Superimposed are  the isocontours  of  the corresponding  resonance
$\Omega - \kappa/2$.  The  dashed line  corresponds to the  isocontour  of a
tenth of the pattern of the wave. Note that the hotter disk forms a 
Grand Design structure involving more eccentric orbits. }
\endCaption
\endfigure

\subsubsection{Lopsided $m=1$  modes}
The algorithm  presented in section~3 is  in principle inadequate to address
the  growth of $m=1$ perturbations which  for  purely self gravitating disks
are forbidden since  the perturbation does  not conserve momentum. Following
Zang\cite{Zang}, it   is assumed  here that  the  disk  is embedded  in some
sufficiently  massive halo to  compensate its  infinitesimal  centre of mass
shift.
\Fig{densityResponse1} gives       the $m=1$    modal      response of   the
isochrone$/m_{10,11}$  disks while   the  corresponding pattern speeds   and
growth rates  are given in   \Tab{growthRateM=1}.  These modes  have smaller
growth rates than their  bi-symmetric counterparts and are practically  more
difficult to isolate because they are close to other modes which grow almost
as  fast; the nyquist  diagrams show many large  loops, all of which must be
properly sampled  to avoid   spurious   contours encircling  the  origin  as
illustrated in
\Fig{m1NyquistDiagram}. 
Indeed, when investigating weaker and weaker  modes the amount of looping in
the corresponding nyquist diagrams  should increase since modes always occur
in pairs ($\omega \pm \eta$) so that when $\eta$ is small one is always in a
r\'egime corresponding at least to two (a growing and a decaying) modes.

 The physical mechanism leading to the appearance of these modes needs to be
clarified, but presents little practical interest when their growth rates do
not correspond to the fastest growing  mode.  The detailed analysis of $m=1$
modes for   the  isochrone disk   is   therefore delayed  until   models  of
distribution functions which display weaker $m=2$ modes are designed.

\midfigure{m1NyquistDiagram}
\vskip 6.5cm
\special{voffset=-20 hoffset=0 hscale=32 vscale=32 psfile=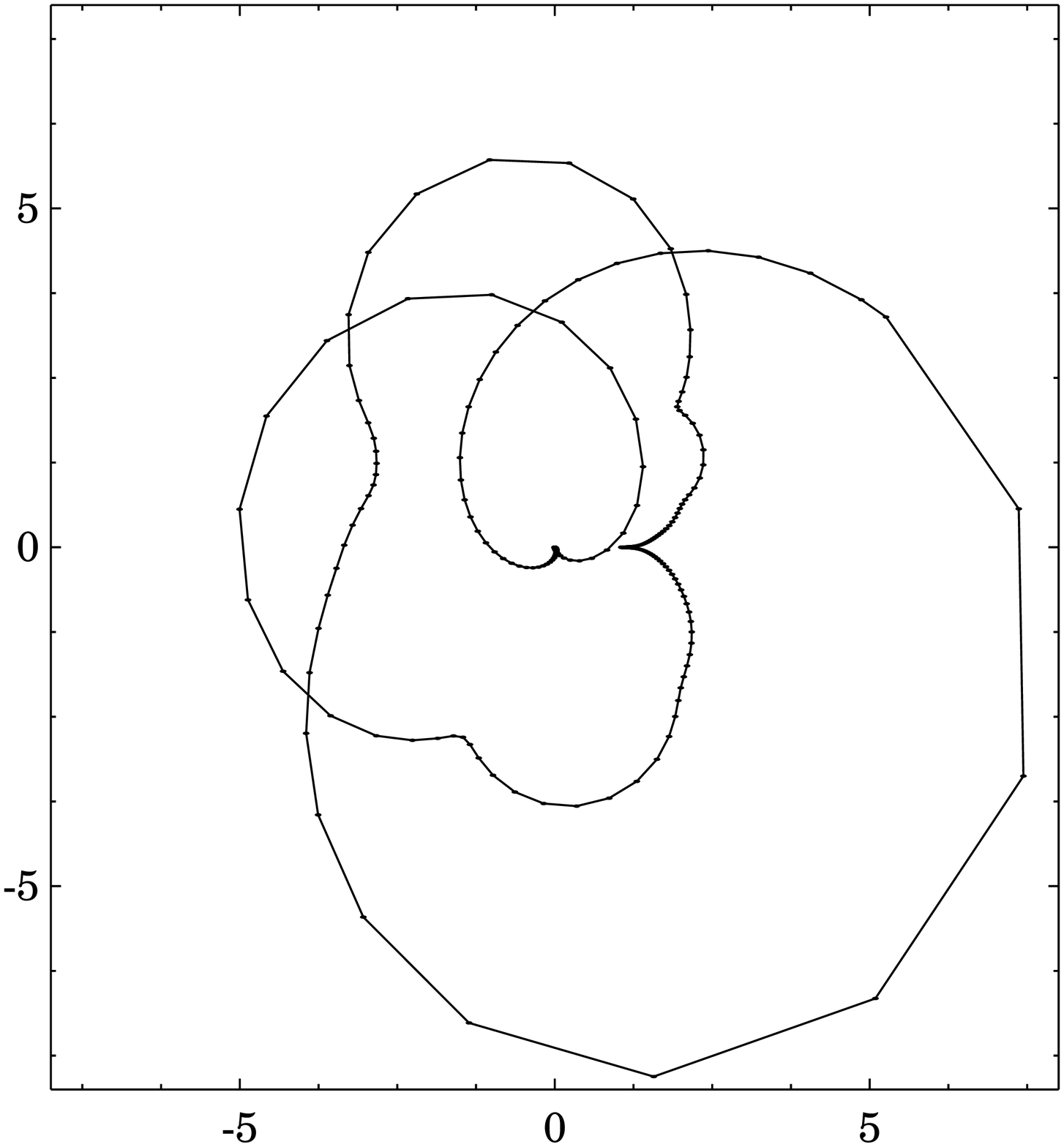}
\special{voffset=-20 hoffset=220 hscale=32 vscale=32 psfile=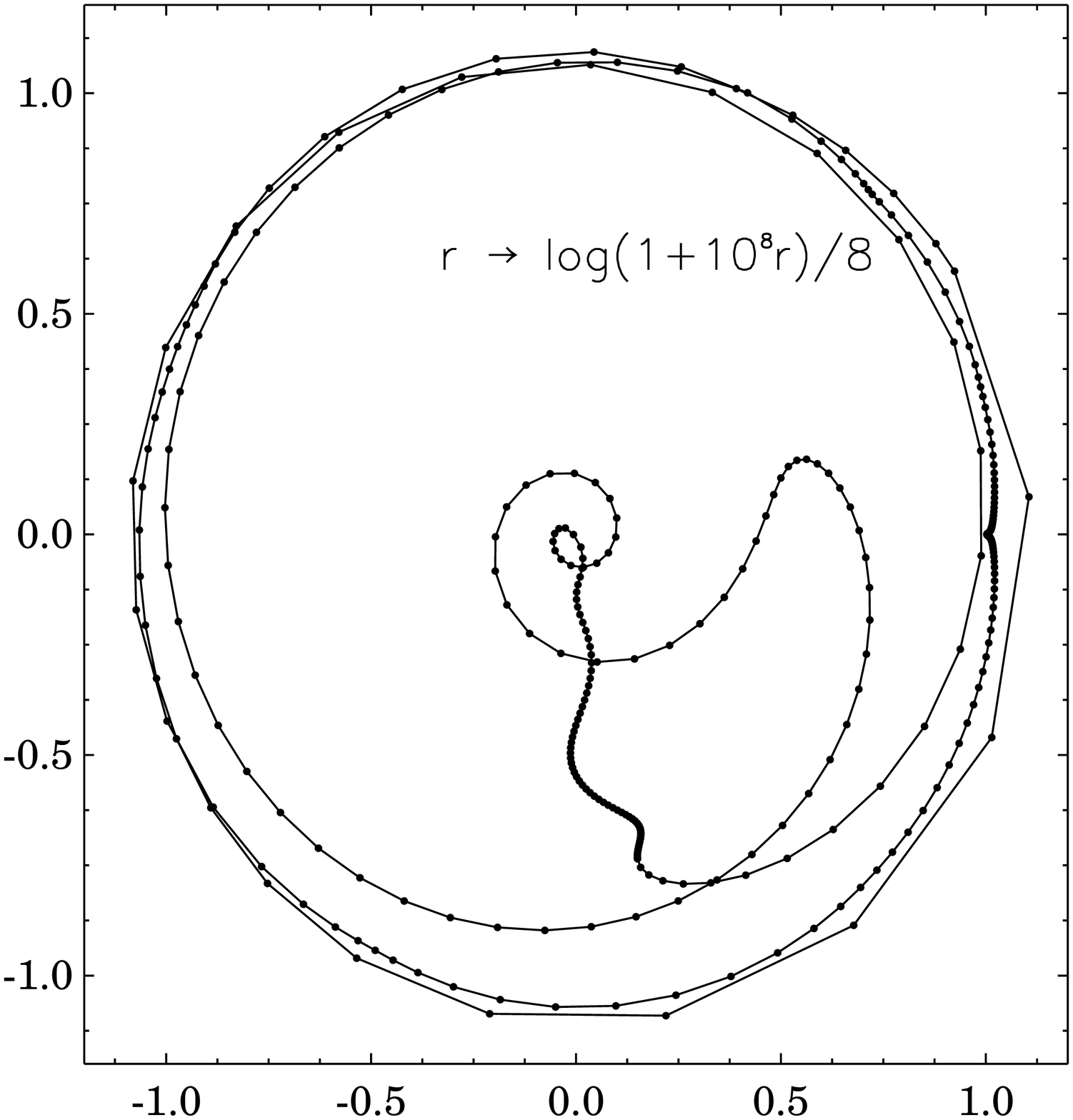}
\Caption{Nyquist diagram for the second $m = 1$ modes of the 
 an isochrone/12 disk. There are more large loops than are seen for the
 bi-symmetric modes, and also more structure at small scales: on the
 right, the origin is shown enlarged under the mapping 
$r \rightarrow (1 + 10^8 r) / 8$. All the space within the innermost of the
 three large concentric loops corresponds to the tiny cusp at the
 centre of the left hand panel.}
\endCaption
\endfigure

\figure{densityResponse1}
\vskip 7cm
\special{voffset=-20 hoffset=0 hscale=40 vscale=40 psfile=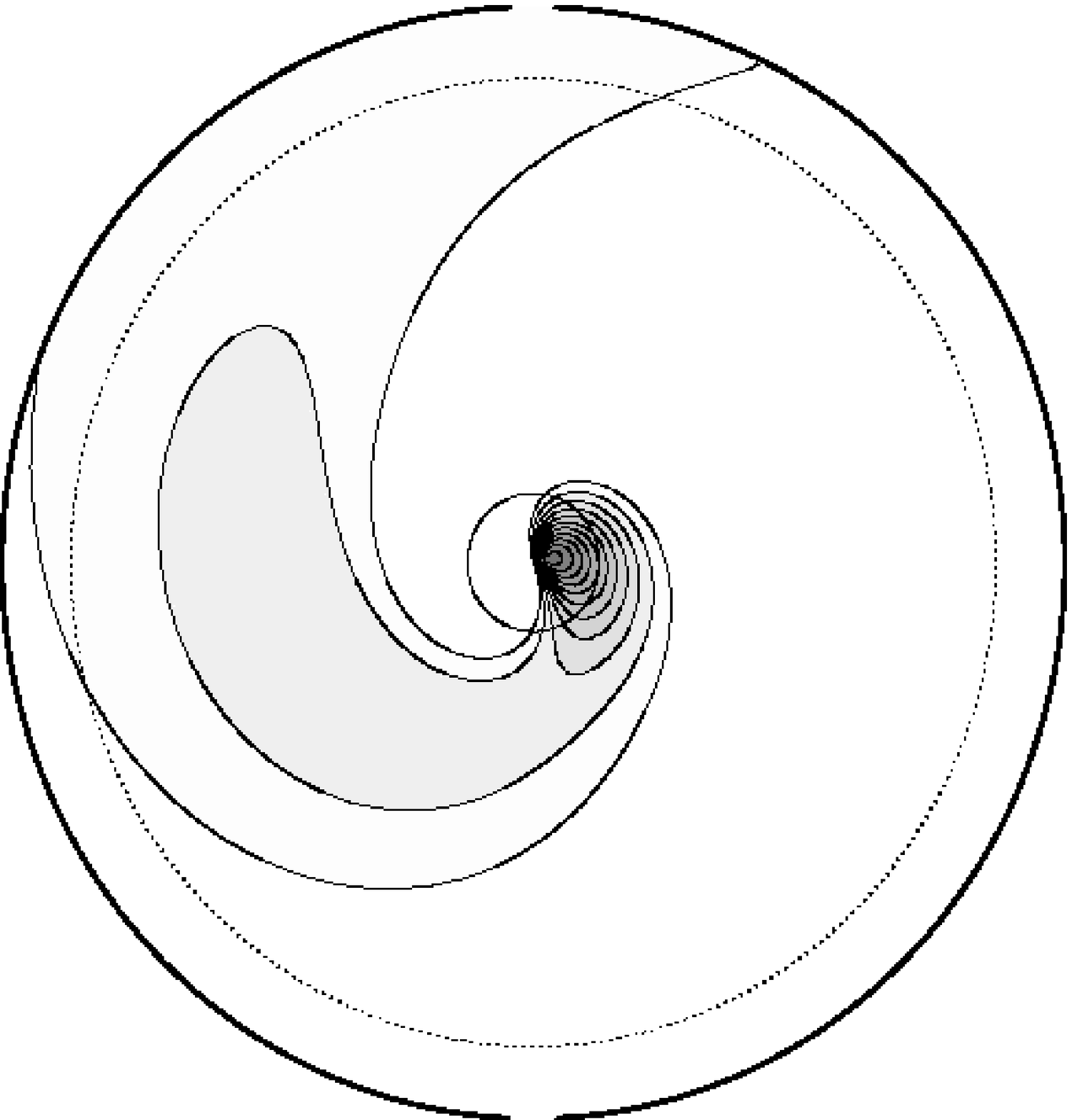}
\special{voffset=-20 hoffset= 200 hscale=40 vscale=40 psfile=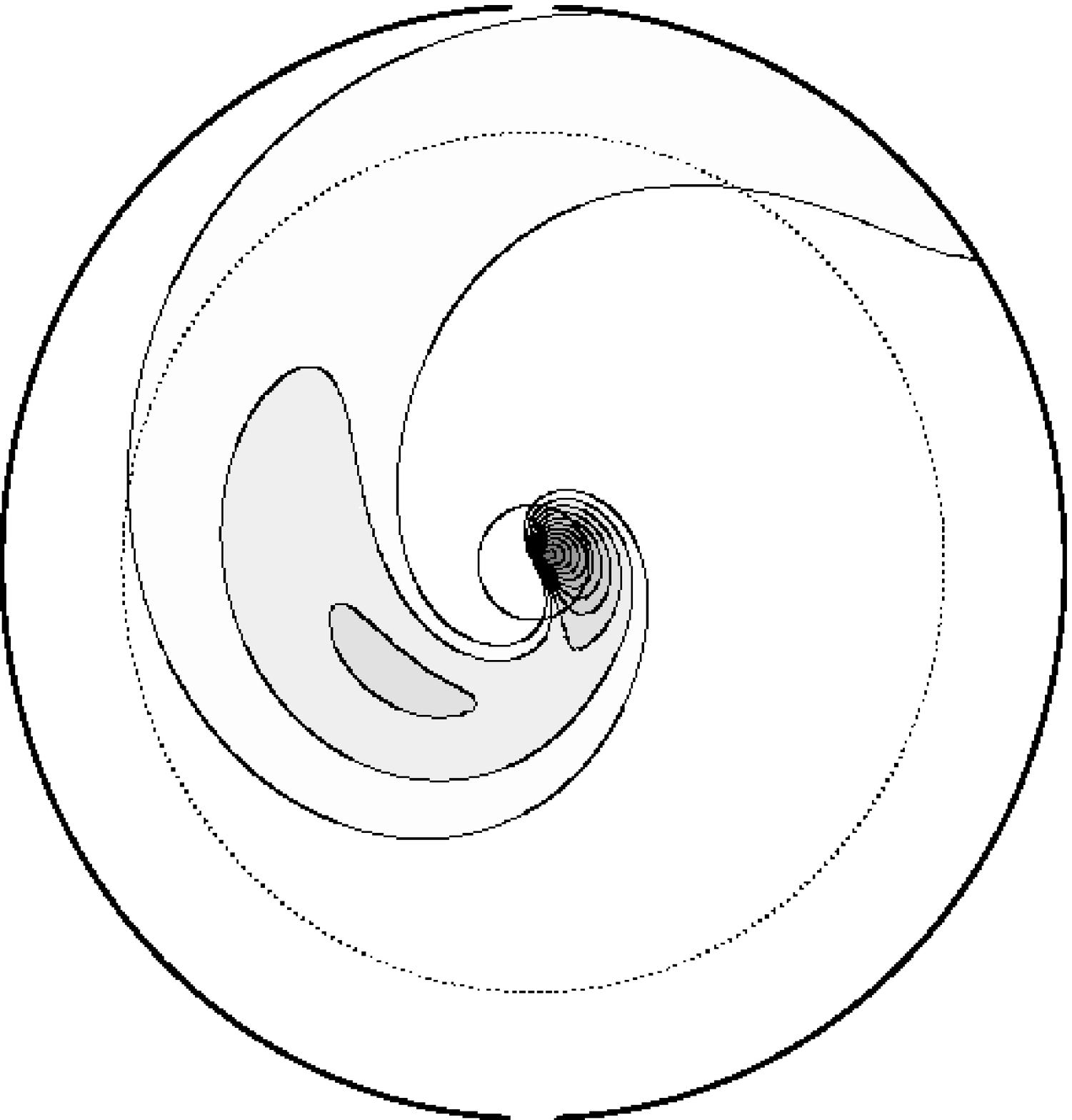}
\Caption
{the  $m=1$ density response corresponding to  the growing mode of the
isochrone$/10$ and isochrone$/11$  model.  The lopsided  mode,  which has  a
smaller  growth rate, has a much   more centrally concentrated response than
its bi-symmetric counterpart.  }
\endCaption
\endfigure

\figure{densityResponse1-2}
\vskip 7cm
\special{voffset=-20 hoffset= 20 hscale=40 vscale=40 psfile=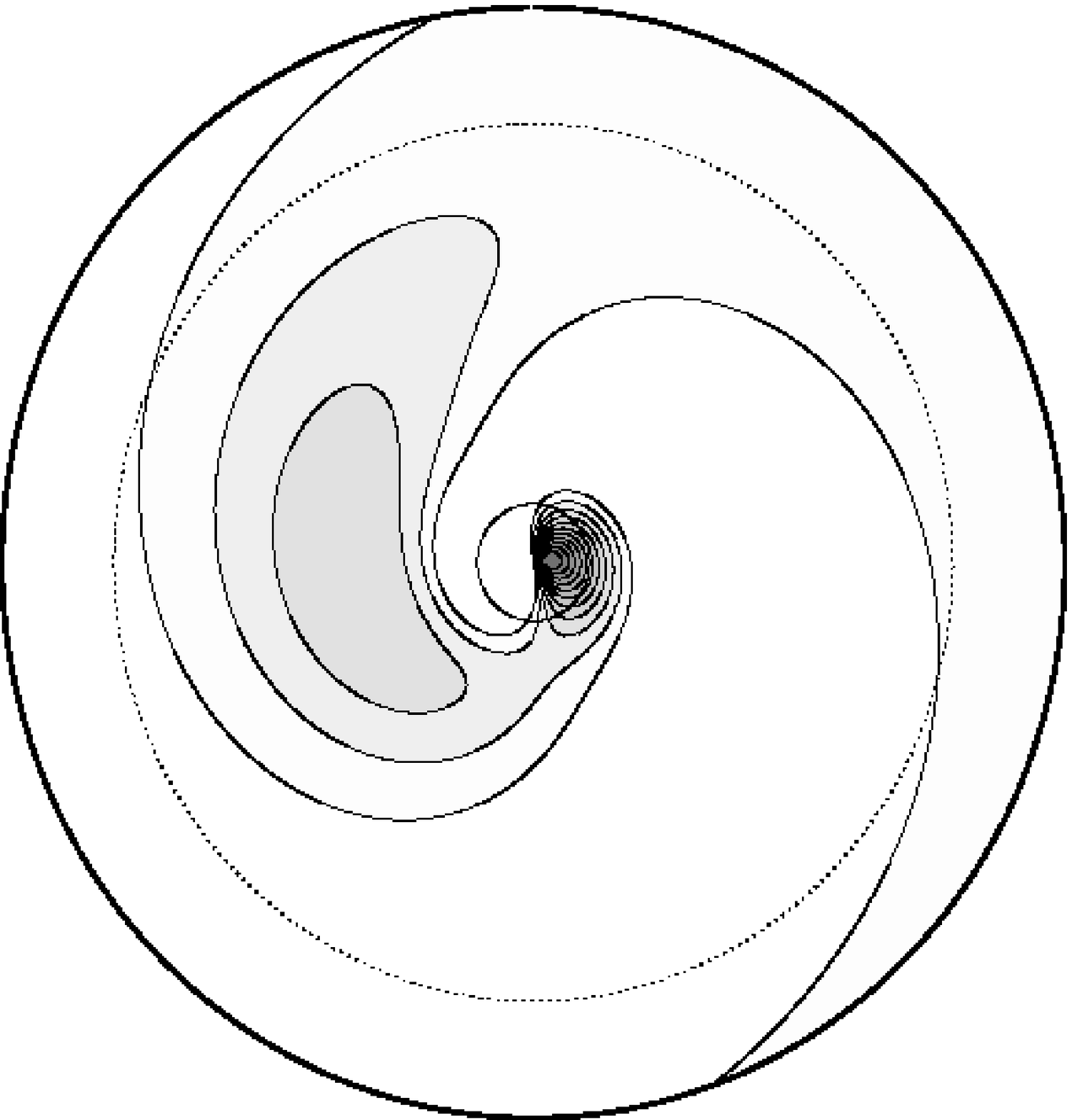}
\Caption
{the $m=1$ density response corresponding to the  second faster growing mode
($\Omega_p+i \eta =  0.18+i\,0.075$) of the  isochrone$/12$ model. The number of
radial  nodes  in  the response  has   increased  by  one compared  to   the
corresponding faster mode illustrated on
\Fig{densityResponse1} for an isochrone$/11$ disk. }
\endCaption
\endfigure

\table{growthRateM=1}
\caption{first and second growth rates and pattern speed of the $m=1$ isochrone/$m_{K}$ model.}
\intablelist{glop.}
\singlespaced
\ruledtable
\multispan3\hfil The  isochrone/$m_K$ model: lopsided mode \hfil\CR
 Model  \dbl pattern speed |  growth rate  \cr 
~~~~9  \dbl  0.135 | 0.0325 \cr
~~~10  \dbl  0.17 | 0.065 \cr
~~~11 \dbl 0.20 | 0.095 \cr 
~~~12 \dbl 0.23 | 0.125
\endruledtable
\endtable

\subsection{Application to observed disks}

The procedure described in section~3  makes no assumption  on the nature  of
the  distribution function or the potential  of the disk.  In particular, it
is  well adapted to  distribution  functions recovered from measured  disks,
where the radial  derivative of the  potential follows from the HI  rotation
curve of the  disk while the distribution function  itself is  inverted from
line of  sight velocity   profiles.  The technique   was  tested  on  tables
generated from \Eq{hunterdf}.  The agreement between the ``theoretical'' and
``measured'' ({\it  i.e.} derived from  a discretized representation  of the
distribution) growth rates was found to be in  better than one percent.  One
peculiarity in  this context lies  in  the finite  difference chosen in  the
numerical derivatives of the  distribution function with  respect to $J$ and
$h$.  While applying this analysis to observed data,  special care should be
taken in handling measurements relative to the core of the galaxy, since the
relative  fraction  of  counter   rotating stars  plays   a crucial  role in
determining the growth rates of the disk,  as pointed out already by Kalnajs
and illustrated in
\Fig{fretplot}.  This  appears clearly when  recalling that counter-rotating
stars add up  to  an effective azimuthal  pressure  in the inner core  which
therefore  prevents  the self  gravity of  the   disk to build  up  by orbit
alignment.  A non parametric inversion  technique has been devised by Pichon
\&  Thi\'ebaut   (1997)\cite{Pichon3}  to   recover the   best  distribution
accounting  for all the measured kinematics  while handling specifically the
counter-rotating stars.

\midfigure{fretplot}
\vskip 6.5cm
\special{voffset=-20 hoffset=100 hscale=40 vscale=60 psfile=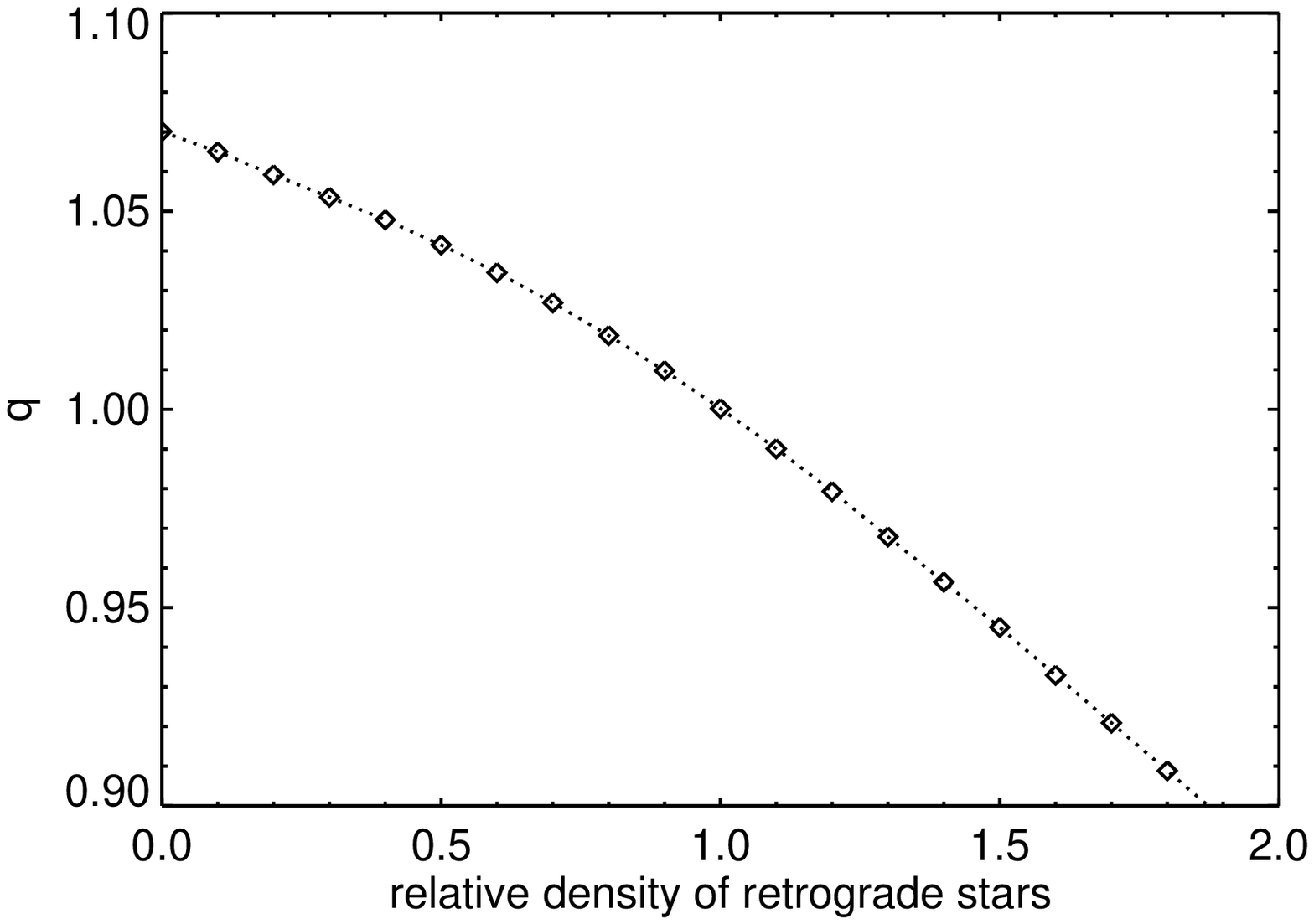}
\Caption
{  
the effect on q of artificially modifying
the density of retrograde stars in Kalnajs' isochrone/9 distribution
function. The retrograde part of the distribution function has been 
multiplied by the factor on the ordinate, keeping the prograde stars
the same. Although in Kalnajs' model stars on retrograde orbits only 
account for a few  percent of the total mass, they play a
significant role in stabilising the disk.}
\endCaption
\endfigure

\section{ Conclusion \& prospects}

A numerical  investigation of linear  stability of round  galactic disks has
proven  successful in recovering known  growth rates for  the isochrone disk
and the Kuzmin Toomre disks.   The method is fast and  versatile, and can be
applied to realistic disks with  arbitrary density and velocity profiles and
relative ``halo'' support.  Unstable  bi-symmetric  growing modes for   {\sl
new}  equilibria,  and  the fastest  growing    lopsided modes for  Kalnajs'
distribution   function have also been   isolated  to demonstrate the code's
versatility.  The authors  are currently applying the method  in a  study of
the   stability to   families  of   disks  parametrised by   their  relative
temperature, compactness, fraction of counterrotating stars and halo support
in order to  identify the  orbits responsible  for the instability,  and  to
probe  the intrinsic  (orbital    or wave-like) nature   of  the  $m=2$  bar
instability. The nature of lopsided $m=1$ instabilities in disk galaxies is
also under investigation.  The  implementation of linear  stability analysis
for observed   galactic disk,  yielding a  direct   relationship between the
growth rate of the instability, the  measured kinematical characteristics of
the disk and the relative mass in the halo  is also being investigated.  The
requirement of marginal stability  will provide an  estimate of  the minimum
halo mass.   All ingredients will then  be in place  to probe  the amount of
dark matter required to stabilise observed galactic disks.

\vskip 0.25cm

{\sl Acknowledgements}~\vskip 0.25cm~{\it 
CP wishes to thank J.~Collett, D.~Lynden-Bell, S.~Tremaine, J.F.~Sygnet
and  O.~Gerhard for  useful conversations.   Many  thanks  to
the referee, J. Sellwood, for positive criticism, to  J.  Binney for
reading the manuscript and  to P. Englemaier for   his help with {\tt  sm}.
Funding from the Swiss NF and computer resources from the IAP are gratefully
acknowledged.  }


\nosechead{References}
\ListReferences
\vfill
\eject

\bye